\documentclass[12pt,aps,prd,floatfix,nofootinbib,a4paper,superscriptaddress]{revtex4-2}
\pdfoutput=1

\usepackage{amsmath, amssymb, amsfonts, amsthm, latexsym, epsfig, mathrsfs, xcolor, bbm, slashed, braket, thmtools,soul}

\usepackage[all]{xy}

\usepackage[inline]{enumitem}

\usepackage{setspace}
\usepackage[marginal, multiple]{footmisc}

\usepackage[T1]{fontenc}
\usepackage[utf8]{inputenc}
\usepackage{lmodern}

\usepackage[colorlinks, allcolors=blue!70!black, linktocpage]{hyperref}

\definecolor{indigo(dye)}{rgb}{0.0, 0.25, 0.42}

\numberwithin{equation}{section}

\usepackage{cleveref}

\usepackage{microtype}

\usepackage{floatrow}
\let\OLDtableofcontents\tableofcontents
\renewcommand\tableofcontents[1]{%
    {\baselineskip 0.5ex %
	\OLDtableofcontents{#1}}%
}

\let\OLDthebibliography\thebibliography
\renewcommand\thebibliography[1]{%
	\setstretch{1.079} 
	\OLDthebibliography{#1}%
	\small %
	\setlength{\itemsep}{0.2\baselineskip} 
}

\let\OLDfootnote\footnote
\renewcommand\footnote[1]{%
	\setlength{\footnotesep}{0.75\baselineskip}%
	{\footnotesize \OLDfootnote{#1}}%
}

\setlist[enumerate]{noitemsep, label=(\arabic*), ref=(\arabic*)}

\newlist{condlist}{enumerate}{2}
\setlist[condlist,1]{noitemsep, topsep=0pt, label=(\arabic*), ref=(\arabic*)}
\setlist[condlist,2]{noitemsep, label=(\alph*), ref=(\arabic{condlisti}.\alph*)}
\crefname{condlisti}{condition}{conditions}
\crefname{condlistii}{condition}{conditions}

\newlist{propertylist}{enumerate}{1}
\setlist[propertylist,1]{noitemsep, topsep=0pt, label=(\arabic*), ref=(\arabic*)}
\crefname{propertylisti}{Property}{Properties}

\renewcommand\thesection{\arabic{section}}
\renewcommand\thesubsection{\arabic{subsection}}

\makeatletter
\def\p@subsection{\thesection.}
\def\p@subsubsection{\thesection.\thesubsection.}
\makeatother 

\theoremstyle{plain}
\newtheorem{thm}{Theorem}
\newtheorem{lemma}{Lemma}[section]
\newtheorem{prop}{Proposition}[section]
\newtheorem{corollary}{Corollary}[section]

\theoremstyle{definition}

\newtheorem{ass}{Assumption}

\declaretheorem[style=remark,qed=$\scriptstyle{\blacksquare}$,numberwithin=section]{remark} 

\creflabelformat{equation}{#2#1#3}

\crefname{section}{\S}{\S}
\crefname{appendix}{Appendix}{Appendices}
\crefname{figure}{Fig.}{Figs.}
\crefname{table}{Table}{Tables}

\crefname{definition}{Def.}{Defs.}
\crefname{prop}{Prop.}{Props.}
\crefname{lemma}{Lemma}{Lemmas}
\crefname{corollary}{Cor.}{Cors.}
\crefname{thm}{Theorem}{Theorems}
\crefname{remark}{Remark}{Remarks}

\crefname{ass}{Assumptions}{Assumptions}
\crefname{property}{Properties}{Properties}

\newcommand{\be}{\begin{equation}\begin{aligned}}
\newcommand{\ee}{\end{aligned}\end{equation}}

\newcommand{\lb}{\left}
\newcommand{\rb}{\right}

\newcommand{\mc}{\mathcal}

\newcommand{\ms}{\mathscr}
\newcommand{\mf}{\mathfrak}
\newcommand{\bb}{\mathbb}

\newcommand{\eqsp}{\, ,\quad} 

\newcommand{\Lie}{\pounds} 
\newcommand{\defn}{\mathrel{\mathop:}=} 

\newcommand{\norm}[1]{\lb\Vert\, #1 \,\rb\Vert}		

\newcommand{\op}[1]{\boldsymbol{#1}}
\renewcommand{\1}{\op{1}}

\newcommand{\inn}{\textrm{in}}

\newcommand{\out}{\textrm{out}}

\newcommand{\Alg}{\ms{A}}

\newcommand{\ground}{\varnothing}

\let\oldsetminus\setminus
\renewcommand{\setminus}{\!\oldsetminus\!} 

\newcommand{\pd}[2][]{\frac{\partial #1}{\partial #2}} 

\let\oldint\int
\renewcommand{\int}{\oldint\limits}

\let\oldlim\lim
\renewcommand{\lim}{\oldlim\limits}

\renewcommand{\bar}{\overline}

\newcommand{\grp}[1]{{\tt #1}}

\newcommand{\scri}{\ms I}

\newcommand{\Hilb}{\mathscr{H}}

\newcommand{\antiHilb}{%
\hspace{4pt} 
  \vbox{%
    \hrule height 0.5pt
    \kern0.25ex
    \hbox{%
      \kern-0.3em
      \ifmmode\Hilb\else\ensuremath{\Hilb}\fi
      \kern0em
    }
  }
}

\newcommand{\Fock}{\mathscr{F}}

\newcommand{\EM}{\textrm{EM}}

\newcommand{\nfrac}[2]{{{}^#1\!\!/\!_#2}}

\newcommand{\half}{\nfrac{1}{2}}

\newcommand{\orbit}{\ms{O}}

\DeclareMathOperator{\Sym}{\text{\rm Sym}}

\renewcommand{\Re}{{\rm Re\,}}
\renewcommand{\Im}{{\rm Im\,}}
\newcommand{\abs}[1]{\lb\vert\, #1 \,\rb\vert}		

\newcommand{\s}{\omega}

\begin{document}

\setstretch{1.2}


\title{Infrared finite scattering theory: Amplitudes and soft theorems}

\author{Kartik Prabhu}
\email{kartikprabhu@rri.res.in}
\affiliation{Raman Research Institute, Sadashivanagar, Bengaluru 560080, India.}

\author{Gautam Satishchandran}
\email{gautam.satish@princeton.edu}
\affiliation{Princeton Gravity Initiative, Princeton University, Jadwin Hall, Washington Road, Princeton NJ 08544, USA}

\begin{abstract}
Any non-trivial scattering with massless fields in four spacetime dimensions will generically produce an out-state with memory. Scattering with any massless fields violates the standard assumption of asymptotic completeness --- that all ``in'' and ``out'' states lie in the standard (zero memory) Fock space --- and therefore leads to infrared (IR) divergences in the standard $S$-matrix amplitudes.  In this paper we define an infrared finite scattering theory which assumes only (1) the existence of in/out algebras and (2) that Heisenberg evolution is an automorphism of these algebras. The resulting ``superscattering'' map $\$$ allows for transitions between different in/out memory states and agrees with the standard $S$-matrix when it is defined. We construct $\$$-amplitudes by defining (3) a ``generalized asymptotic completeness'' which accommodates states with memory in the space of asymptotic states and (4) a complete basis of improper states which generalize the usual $n$-particle momentum basis to account for states with memory. Using only general properties of $\$$, we prove an analog of the Weinberg soft theorems in quantum gravity and QED which imply that all $\$$-amplitudes are well-defined in the infrared. We comment on how one must generalize this framework to consider $\$$-amplitudes for theories with collinear divergences (e.g., massless QED and Yang-Mills theories). 
\end{abstract}

\maketitle
\newpage 
\tableofcontents

\section{Introduction}\label{sec:intro}

The basic assumptions of conventional scattering theory, as developed by Lehmann-Symanzik-Zimmermann (LSZ) \cite{LSZ} as well as Haag and Ruelle \cite{Haag,Ruelle1962}, are
\begin{enumerate*}
    \item the existence of an ``in/out'' algebra of asymptotic fields\footnote{The asymptotic algebra is generally taken to be defined as a (weak) limit of the bulk algebra \cite{LSZ,Haag,Ruelle1962}. We will not concern ourselves here with the precise limiting relationship between the bulk and asymptotic algebras and merely assume that such in/out algebras exist.},
    \item that the Heisenberg evolution is an automorphism on these algebras (i.e. maps the ``in'' algebra to the ``out'' algebra and vice-versa), and 
    \item that all ``in/out'' quantum states at asymptotically early and late times lie in the standard Fock space $\Fock_{0}$.
\end{enumerate*}
This third assumption is known as {\em asymptotic completeness}. Given these assumptions, the scattering can be represented as a unitary operator $\op{S}:\Fock_{0}\to \Fock_{0}$ on the standard Fock space known as the $S$-matrix. The properties of this map can be obtained by computing \(S\)-matrix amplitudes given by the components of $\op{S}$ in a basis of $\Fock_{0}$ \be\label{eq:amplitude}
    \braket{p_1^\out,\ldots,p_n^\out | \op S | p_1^\inn,\ldots,p_m^\inn}
\ee
where the standard basis is the ``$n$-particle momentum basis'' of $\Fock_{0}$. This approach has been extremely successful in providing a satisfactory formulation of scattering theory for massive fields \cite{Haag,Ruelle1962,LSZ}. However, for theories involving interacting, massless fields (e.g., QED, Yang-Mills or quantum gravity), the usual $S$-matrix formulation of scattering suffers from infrared divergences. Indeed, by the Weinberg soft theorems \cite{Weinberg:1965}, the \(S\)-matrix amplitudes given by \cref{eq:amplitude} are divergent at low frequencies which renders the out-state non-normalizable in $\Fock_{0}$ and, consequently, the standard $S$-matrix cannot be defined.\footnote{In $d$ spacetime dimensions where the dimensions are non-compact, the matrix elements of $\op{S}$ have a pole at zero frequency. This pole is integrable in $d>4$ dimensions, but is a non-integrable singularity in $d=4$ (see, e.g. \cite{Kapec:2015vwa,Ferko:2021bym} and \S~G. of \cite{Satishchandran_2019}). This is related to the fact that the expected number of outgoing radiative quanta generically diverges in four dimensions but is finite in higher dimensions.} These divergences imply that one of the above three assumptions must be incorrect for theories with massless particles. Since the ``in'' and ``out'' algebras and the Heisenberg evolution of the fields are well-defined, the key assumption that is violated is asymptotic completeness. In this paper we define a scattering theory based only on the Heisenberg evolution between the in/out algebras and --- by generalizing the assumption of asymptotic completeness --- we will define scattering amplitudes that are normalizable in the infrared.

This failure of asymptotic completeness and the resulting infrared divergences in the standard \(S\)-matrix are 
directly related to the so-called \emph{memory effect}, where the leading-order massless field at late times does not go back to its original value at early times. In general relativity, the radiative field is represented by the shear tensor $\sigma_{AB}$ at null infinity, whose time-derivative gives the Bondi news tensor $N_{AB}=2\partial_{u}\sigma_{AB}$ characterizing gravitational radiation. The memory $\Delta_{AB}(\hat x)$ captures the low-frequency behavior of this radiative field
\be\label{eq:mem}
\Delta_{AB}(\hat x) = \frac{1}{2} \int_{\bb R}du~ N_{AB}(u,\hat x) = \sigma_{AB}(u=+\infty,\hat x) - \sigma_{AB}(u=-\infty, \hat x) 
\ee
Here \(u\) is the retarded (or advanced) time coordinate on future (or past) null infinity, \(\hat x\) denotes a point on the \(2\)-sphere and the indices \(A,B,\ldots\) are abstract tensor indices on the \(2\)-sphere. Classically, even if the initial data in the far past has a vanishing memory, the outgoing radiative field will generically have non-vanishing memory \cite{Zeldovich:1974gvh,PhysRevLett.67.1486,Wiseman:1991ss,Bieri_2013EM}. In the quantum theory, a non-vanishing memory implies a divergence in the expected number of radiative quanta at low frequencies, resulting in a corresponding divergence in the standard Fock space norm of states with memory --- all states in $\Fock_{0}$ have vanishing memory \cite{narain81,Ashtekar:1981,asymp-quant}. Thus, scattering with any massless fields {\em violates} the standard assumption of asymptotic completeness. This is the fundamental origin of the infrared divergences one encounters in the standard \(S\)-matrix amplitudes. 

If one is interested in computing cross-sections relevant for collider experiments, an infrared cut-off can be imposed based on the scattering timescales and detector resolution, and ``inclusive cross-sections'' for the scattering of ``hard'' particles can be computed \cite{Bloch_1937,Yennie:1961ad,Weinberg:1965,Frye_2018,Hannesdottir:2019umk,Hannesdottir_2019}. However, on longer timescales, the production of soft quanta becomes physically significant, e.g., leading to complete decoherence of ``hard'' particles at late times \cite{Carney_2017,Semenoff:2018,Semenoff_2019,Danielson_2022,Danielson:2022sga,Gralla:2023oya,Danielson:2024yru}. More significantly, this approach is certainly not satisfactory if one wants to view the \(S\)-matrix as a fundamental object in QFT and quantum gravity --- as is the case in any holographic formulation of quantum gravity (see, e.g., \cite{Strominger:2017zoo,Pasterski:2021raf,Donnay:2023mrd,Pasterski:2023ikd} and references therein). Therefore, to account for all observables --- including low-frequency observables like the memory --- it is essential to formulate a well-defined, infrared finite scattering theory.

To obtain such an infrared finite scattering theory, one must include sufficiently many states with non-zero memory. Such states with memory $\Delta_{AB}(\hat x)$ lie in a Fock space $\Fock_{\Delta}$ which is unitarily inequivalent to the standard Fock space \(\Fock_0\) \cite{asymp-quant}. As emphasized in \cite{PSW-IR,memory-orbits}, in order for the memory states to have finite energy, the allowed memories must be at least square-integrable on the $2$-sphere. Indeed, it will be proven in \cref{sec:amplitudes} that one cannot restrict to a smaller class of, say, smooth memories. Since these Fock spaces are labeled by the tensor field $\Delta_{AB}(\hat x)$, there are uncountably many inequivalent Fock representations, making it impossible to contain them all in a single, separable Hilbert space \cite{PSW-IR}. Additionally, as the memory is not conserved, any arbitrary separable subspace of states will not scatter into itself under evolution.

In QED, where the analogous memory Fock representations $\Fock^\EM_{\Delta}$ are labeled by the electromagnetic memory $\Delta^\EM_{A}(\hat x)$, Faddeev and Kulish \cite{Kulish:1970ut} (based on the earlier work of \cite{Dollard,Chung_1965,Greco:1967zza}) successfully assembled these states together with states of the charged field to obtain a Hilbert space of physical states. This construction relies on the existence of conserved charges associated with the generators of the infinite-dimensional, asymptotic symmetry group --- the group of ``large gauge transformations'' at infinity \cite{KP-EM-match}. The Faddeev-Kulish Hilbert spaces are constructed by correlating the massive charged field states\footnote{As emphasized in \cite{Dollard,Dybalski:2017mip}, the charged fields themselves must also be appropriately ``dressed'' with their asymptotic Coulomb fields.} with states of electromagnetic radiation with the appropriate memory to construct eigenstates of these conserved charges (see e.g, \cite{PSW-IR}). Due to the conservation of these ``large gauge'' charges, the Faddeev-Kulish Hilbert space ``scatters into itself'' under evolution; see also \cite{Chen_2009_I,Chen_2009_II}. Thus, an $S$-matrix can be defined on the Faddeev-Kulish Hilbert space \cite{Gabai:2016kuf,Duch:2019wpf}.  While this construction yields a suitable space of physical states in massive QED, it was shown in \cite{PSW-IR} that it cannot be extended to QED with massless charged fields, Yang-Mills theories, or quantum gravity. In the gravitational case, the relevant symmetry group is the infinite-dimensional Bondi-van der Burg-Metzner-Sachs (BMS) group and the analogous Faddeev-Kulish construction would be to construct eigenstates of the conserved\footnote{Conservation of the BMS charges was proven in \cite{CE,Henneaux:2018gfi,KP-EM-match,KP-GR-match,Mohamed:2021rfg,Mohamed:2023jwv}} supermomentum given by
\begin{equation}
\label{eq:Qi0f}
\mc{Q}_{i^{0}}(f) = \frac{1}{8\pi}\Delta(f) + \mathcal{J}(f)
\end{equation}
where $f(\hat{x})$ is a function on the $2$-sphere and first term on the right hand side of \cref{eq:Qi0f} is the ``null memory'' \cite{Christ_nonlin_mem} and the second term is determined by the memory
\begin{equation}
\mc{J}(f) \defn -\frac{1}{32\pi} \int_{-\infty}^{\infty}du \int_{\bb{S}^{2}}d^{2}\hat{x}  f N_{AB}N^{AB} ~\textrm{ and }~ \Delta(f) \defn \int_{\bb{S}^{2}}d^{2}\hat{x}~\Delta_{AB}\mathscr{D}^{A}\mathscr{D}^{B}f
\end{equation}
where $d^{2}\hat{x}$ is the area element on the sphere and $\ms{D}_{A}$ is an angular derivative. We note that if $f$ is a linear combination of $\ell =0,1$ spherical harmonics then $\Delta(f\vert_{\ell=0,1})$ vanishes and $\mc{Q}_{i^{0}}(f\vert_{\ell = 0,1})$ corresponds to the ADM mass and momentum. However, in quantum gravity, it was proven in  \cite{PSW-IR} that the only physical eigenstate of the supertranslation charges is the vacuum state.\footnote{For linearized gravity the analogue of the Faddeev-Kulish Hilbert spaces can be constructed \cite{Choi:2017bna,Akhizer_1965} with both massive and massless sources.} Consequently, there is no preferred Hilbert space for scattering in general quantum field theories and in quantum gravity, and an \(S\)-matrix cannot be defined.

Nevertheless, the in/out algebras are well defined, as is the Heisenberg evolution of the fields. The key point is that these properties alone are sufficient for obtaining a well-defined scattering theory (see \cref{subsec:IRfinitegencomp}). Given an in-state $\Psi_{\inn}$ --- which may be an element of $\Fock_{0}$ --- on the in-algebra, the correlation functions of the ``out'' state $\Psi_{\out}$ --- which will generally {\em not} be an element of $\Fock_{0}$ --- can be obtained using the Heisenberg equations of motion (see e.g., \cite{YangFeld_1950,Kallen:1950uha,Haag:1955ev,Hollands:2002ux,Duetsch:2004dd,Hollands:2014eia} for the formulation of scattering in ``Heisenberg picture''). Thus, to any order in perturbation theory, the correlation functions of any out-observable can be computed. This procedure is in the spirit of the ``in-in formalism''  \cite{Keldysh:1964ud,Kadanoff-Baym,Weinberg:2005vy} and defines a ``superscattering''\footnote{The terminology of a ``superscattering'' map and the notation $\$$ is adopted from Hawking \cite{Hawking_1976} who was concerned with a very different generalization of quantum field theory to address issues which arise in ``information loss''. We shall not be concerned with such issues (see property \ref{S3}), however we are generalizing the standard framework of scattering theory to accommodate infrared radiation.} map $\$$ from any specified in-state $\Psi_{\inn}$ to the out-state $\Psi_{\out}$ obtained by evolution
\be
    \Psi_\out = \$\,\Psi_\inn\,.
\ee
We emphasize that, from this definition, the construction of the scattering map $\$$ utilizes only assumptions (1) and (2) above. In particular, we do not attempt to ``presuppose'' which ``out'' Hilbert space that $\Psi_{\out}$ lies within. As explained in \cref{subsec:IRfinitegencomp}, in the case where the scattering can be defined on a single Hilbert space (e.g., for massive fields) then the map $\$$ is equivalent to the standard, unitary $S$-matrix $\op{S}$. Indeed, from its general definition, $\$$ satisfies many of the same fundamental properties of the standard S-matrix (see properties \ref{S1}--\ref{S4}). However, the more ``fine-grained'' details of the map $\$$ are not directly transparent in this framework. The situation is similar to the case of scattering with massive fields where the detailed properties of $\op{S}$ are only made manifest when cast in the form of amplitudes. The goal of this paper is to define {\em superscattering amplitudes} which allow for transitions between states with different memory. We emphasize that the purpose of this paper is to derive properties of this map which follow from the general assumptions highlighted above. We will not, in this paper, attempt to deal with any issues related to practical computations of such amplitudes and merely focus on their general properties. 

To construct superscattering amplitudes we will, for definiteness, primarily restrict to the quantum gravitational case, however the construction proceeds in an analogous manner for other quantum field theories which suffer from infrared divergences due to the emission of soft radiative quanta (see remark \ref{rem:IRfinitesoftphoton}). To achieve this we appropriately generalize the two basic ingredients that go into the definition of the usual (ill-defined) scattering amplitudes given by \cref{eq:amplitude}. The first is that we weaken the assumption (3) above and formulate a {\em generalized asymptotic completeness} which accommodates the fact that the out-state will generically have memory. The second key ingredient is that we must suitably generalize the notion of a ``$n$-particle momentum basis'' of $\Fock_{0}$  to a suitable basis of the Fock spaces $\Fock_{\Delta}$ with memory.  The superscattering amplitude is then defined as the ``matrix elements'' of \(\$\) in this basis. 

As previously mentioned, the usual assumption of asymptotic completeness is that every ``in'' and ``out'' state is an element of the standard Fock space for the massive and massless fields. For example, in the vacuum gravitational case, this would assume that all states lie in $\Fock_{0}$.  This assumption is far too restrictive to accommodate the states that arise in generic scattering with massless fields and we therefore formulate a generalized asymptotic completeness which is precisely stated in assumption \ref{genasympcomp} in \cref{subsec:IRfinitegencomp}. Roughly speaking, assumption \ref{genasympcomp} states that the Fock spaces $\Fock_{\Delta}$ for all (square-integrable) $\Delta_{AB}$ suitably ``span'' the space of states which arise in scattering theory. More precisely, we assume that all states can be expressed in terms of a sum --- or integral\footnote{As emphasized in \cite{memory-orbits}, in order to have well-defined angular momentum, the states must be ``continuously distributed'' in memory.} --- of states in the memory Fock spaces $\Fock_{\Delta}$. This assumption appears to be valid for perturbative scattering of scalar fields, QED with massive charges and quantum gravity where the ``non-Fock'' behavior of the state is entirely characterized by the memory content of the radiation field \cite{Bloch_1937,Weinberg:1965}. However, as we explain in \cref{subsec:colldiv}, it will not be valid for theories with collinear divergences such as massless QED and Yang-Mills theories. Therefore, while the scattering map $\$$ is well-defined for general quantum field theories, it seems that one must further broaden the notion of asymptotic completeness to construct well-defined $\$$-amplitudes in these cases. 

Nevertheless, for theories that satisfy generalized asymptotic completeness, an amplitude can then be defined by choosing a suitable basis on each Fock space $\Fock_{\Delta}$ for all $\Delta$. We now give a summary of the construction of the appropriate generalization of the $n$-particle momentum states $\ket{p_{1}\dots p_{n}}$ given \cref{sec:rigged}. Recall that these momentum states are eigenstates of the energy-momentum --- or, more generally, eigenstates of the supermomentum --- in the standard Fock space $\Fock_{0}$. The analogous basis for states of non-vanishing memory are eigenstates of the supermomentum in $\Fock_{\Delta}$. The construction of such eigenstates is highly non-trivial due to the fact that there is no ``ground state'' of the supermomentum --- a normalizable state with $\mc{Q}_{i^{0}}=0$ ---  in a memory Fock space $\Fock_{\Delta}$ with $\Delta\neq 0$. Nevertheless, in \cref{sec:rigged} we use the ``rigged Hilbert space formalism''\footnote{Another approach advocated in \cite{Reed-Simon} is to use sequences of ``projection valued measures'' to construct the improper eigenstates. For purely practical convenience, we will not be following the viewpoint expressed in the Notes to \S~VII in Reed and Simon \cite{Reed-Simon}: ``\textit{We must emphasize that we regard the spectral theorem as sufficient for any argument where a nonrigorous approach might rely on Dirac notation; thus, we only recommend the abstract rigged space approach to readers with a strong emotional attachment to the Dirac formalism.}''} \cite{Gelfand4} to directly construct this improper basis of eigenstates which we refer to as {\em BMS particle states}, denoted by 
\be
    \ket{\ground;\Delta} \eqsp \ket{p_1,\ldots,p_n; \Delta}.
\ee
Here \(\ket{\ground;\Delta}\) can be considered as a ``BMS \(0\)-particle state'' and \(\ket{p_1,\ldots,p_n;\Delta}\) as a ``BMS \(n\)-particle state'' labeled by \(n\) future-directed, null momenta \(p_1,\ldots,p_n\). These states are improper eigenstates of memory with eigenvalue $\Delta_{AB}$ (see \cref{sec:rigged}) and when the memory vanishes this basis coincides with the basis $\ket{p_{1}\dots p_{n}}$ of $\Fock_{0}$. We will show, in theorem \ref{thm:eigenstates}, that the BMS particles are eigenstates of the supermomentum operator --- denoted as $\op{\mc{Q}}_{i^{0}}(f)$ --- with eigenvalue\footnote{In the rigged Hilbert formalism, the improper states are linear functionals on a dense subspace of the Fock space. Thus, an improper eigenstate satisfies the eigenstate condition in the weak sense (see \cref{subsec:RiggedHilbertSpaces}).}
\begin{equation}
\label{eq:eigenimprop}
\op{\mc{Q}}_{i^{0}}(f)\ket{p_1,\ldots,p_n;\Delta} = \bigg(\sum_{i=0}^n \omega_i f(\hat p_i) + \Delta(f)\bigg)\ket{p_1,\ldots,p_n;\Delta} \quad \textrm{ (on $\Fock_{\Delta}$)}
\end{equation}
where $\omega_{i}$ is the frequency, $\hat{p}_{i}$ is the direction of the future directed null vector $p_{i}$, and this equation is to be understood in the sense of improper eigenstates as explained in \cref{sec:rigged}. In theorem \ref{thm:completeness}, we prove that these eigenstates form a complete basis of $\Fock_{\Delta}$.\footnote{The existence of improper states satisfying \cref{eq:eigenimprop} were previously conjectured in \cite{Strominger:2013jfa,Ashtekar:2018lor}}

Given these two ingredients, the scattering of any in-state $\Psi_{\inn}$ to any out-state $\Psi_{\out}$ can be expressed in terms of superscattering $\$$-amplitudes as ``components'' of the superscattering map in the BMS particle basis: 
\begin{equation}
\label{eq:superamp}
\braket{p_{1}^{\textrm{out}},\dots ,p_{1}^{\textrm{out}};\Delta^{\textrm{out}}|\,\$\,|p_{1}^{\textrm{in}},\dots ,p_{1}^{\textrm{in}};\Delta^{\textrm{in}}}
\end{equation}
which, in contrast to $\op{S}$, allows for transitions between states with different memories. We then establish various properties of the superscattering amplitudes. We first show, using conservation of supermomentum, that any non-trivial $\$$-amplitude must have non-vanishing change in memory between the ``in'' and ``out'' states as originally noted in \cite{Ashtekar:2018lor}. Furthermore, we show that while the memories which arise in scattering theory are not smooth they are square-integrable on the $2$-sphere and therefore have finite energy --- consistent with generalized asymptotic completeness. Finally, in theorem \ref{thm:soft} we prove that the conservation of supermomentum implies a {\em soft theorem} for the $\$$-amplitudes which is analogous to the Weinberg soft graviton theorem. More precisely, we show that the $\$$-amplitudes factorize in the limit as one of frequencies of the BMS particle momenta vanishes. Explicitly, this factorization takes the form 
\be
\label{eq:Weinberg}
    &\lim_{\omega_n^\out \to 0} \braket{p_1^\out,\ldots,p_n^\out; \Delta^\out | \,\$\, | p_1^\inn,\ldots,p_m^\inn; \Delta^\inn} = S^{(0)}_{A_{n}B_{n}} \braket{p_1^\out,\ldots,p_{n-1}^\out; \Delta^\out | \,\$\, | p_1^\inn,\ldots,p_m^\inn; \Delta^\inn}
\ee
where the $A_{n}B_{n}$ indices correspond to tensor indices of the $n$-th BMS particle and the coefficient $S_{AB}^{(0)}$ is given in \cref{eq:S0}. For the case where the incoming memory vanishes (i.e.  $\Delta^{\inn}=0$) the components $S^{(0)}$ of $S_{AB}^{(0)}$ in a normalized basis (defined in \cref{subsec:IRamp}) are is given by 
\begin{equation}
S^{(0)} =- \sqrt{\frac{1}{8\pi^{3} n}}~ \lb( \sum_{i=0}^m \frac{p_i^{\inn, \mu} p_i^{\inn, \nu} \bar\epsilon_{\mu\nu}}{p_i^{\inn, \mu} \hat{p}_n^{\out, \nu} \eta_{\mu\nu}} - \sum_{i=0}^{n-1} \frac{p_i^{\out, \mu} p_i^{\out, \nu} \bar\epsilon_{\mu\nu}}{p_i^{\out, \mu} \hat{p}_n^{\out, \nu} \eta_{\mu\nu}} \rb),
\end{equation}
where $\hat{p}^{\out}_{n}$ is the angular direction of the $n$-th null momentum and \(\bar\epsilon_{\mu\nu}\) is a polarization tensor relative to \(p_n\) (see \cref{subsec:IRamp}). 

These results establish that the infrared properties of $\$$-amplitudes are analogous to that of $\op S$-amplitudes. In what sense are the $\$$-amplitudes infrared finite? We recall that the original amplitude $\braket{p^{\out}_{1}\dots p^{\out}_{n};0|\op{S}|p^{\inn}_{1}\dots p^{\inn}_{n};0}$ is attempting to calculate the amplitude to scatter from an ``in'' zero memory plane wave state to an ``out'' plane wave state with zero memory. For non-trivial scattering, memory is generically produced and so when one attempts to normalize this amplitude in the zero memory Fock space $\Fock_{0}$ one obtains a divergence. In contrast, we prove that the $\$$-amplitude $\braket{p^{\out}_{1}\dots p^{\out}_{n};0|\$||p^{\inn}_{1}\dots p^{\inn}_{n};0}$ identically {\em vanishes} which is the physically correct answer. Furthermore, by eq.~\ref{eq:Weinberg}, a general $\$$-amplitude $\braket{p^{\out}_{1}\dots p^{\out}_{n};\Delta^{\textrm{out}}|\$||p^{\inn}_{1}\dots p^{\inn}_{n};\Delta^{\textrm{in}}}$ is normalizable as an amplitude to scattering map between states in $\Fock_{\Delta^{\inn}}$ to $\Fock_{\Delta^{\out}}$.

The construction of well-defined scattering amplitudes opens the door to a number of lines of inquiry into the more formal aspects of scattering theory. In particular, it would be of interest to prove other ``soft theorems'' associated with e.g., Lorentz charges \cite{Cachazo:2014fwa}, soft gluon theorems \cite{He:2015zea} as well as sub-leading soft theorems \cite{Guevara:2021abz,Strominger:2021mtt}. These soft theorems either involve charges that change the memory of the state --- and so must be dealt with using the methods of \cite{memory-orbits} --- or analyze higher-frequency behavior of the scattering map than the leading soft theorems. Additionally, in the study of standard S-matrix amplitudes, properties such as analyticity and crossing symmetry \cite{Bros:1964iho,Bros:1965kbd,Bros:1972jh} play a crucial role. While the $\$$-amplitudes given by \cref{eq:superamp} appear to depend on in/out-momenta as well as the in/out-memory, we prove in \cref{subsec:IRamp} that the $\$$-amplitude vanishes unless the out-memory is expressed entirely in terms of the in/out-momenta as well as the in-memory. Therefore if we consider the case where $\Delta^{\inn}$ vanishes --- as is usually assumed ---  then \cref{eq:superamp} is entirely a function of the in/out-momenta. Thus, properties of $\$$-amplitudes such as analyticity and crossing symmetry can also be investigated in an analogous manner. Finally, at a more formal level, the construction of the momentum basis of $\Fock_{0}$ is directly tied to Wigner's construction of irreducible representations of the Poincaré group. While not necessary for the construction of a well-defined scattering theory (see remark \ref{rem:bms-particles-mccarthy}), it would be interesting to analyze the relationship between the BMS particles basis of $\Fock_{\Delta}$ and the irreducible representations of the BMS group classified by McCarthy \cite{McCarthy-nucl,McCarthy1,McCarthy2}.

The structure of the rest of this paper is organized as follows. In \cref{sec:asymp-quant} we provide a brief review the asymptotic quantization of general relativity at null infinity, and construct the asymptotic quantization of the algebra of radiative fields together with the BMS charges. In \cref{sec:mem-fock} we review the construction of the memory Fock spaces \(\Fock_\Delta\) and in \cref{sec:coherent-states} we illustrate how the memory vacua can be viewed as a generalization of coherent states. In \cref{sec:rigged} we consider the improper eigenstates of the supermomentum. In \cref{subsec:RiggedHilbertSpaces} we review the rigged Hilbert space formalism and in \cref{sec:bms-particles} we apply this formalism to explicitly obtain the BMS particle states described above. In \cref{sec:amplitudes} we construct and analyze the properties of infrared finite $\$$-amplitudes. In \cref{subsec:IRfinitegencomp}, we define an infrared finite superscattering map $\$$, enumerate the basic properties of this map, and formulate the notion of generalized asymptotic completeness. In \cref{subsec:IRamp} we define the infrared finite $\$$-amplitudes and prove a well-defined, infrared finite soft theorem for these amplitudes. Finally, in \cref{subsec:colldiv}, we consider the case of collinear divergences and illustrate that one must further generalize the notion of asymptotic completeness to formulate well-defined amplitudes in such theories. 

In the appendices, we collect some results on the BMS particles which are not strictly necessary for the arguments in the main text. In appendix \ref{sec:approx-eigenstates}, we show that the BMS particle states can be approximated by a sequence of coherent states with memory $\Delta_{AB}$ which limit to an improper state with support at only zero frequency. In appendix \ref{sec:inp-mem} we obtain an explicit expression for the inner product on the memory Fock spaces in terms of the BMS particle states. 

\section{Asymptotic quantization of general relativity}
\label{sec:asymp-quant}

In this section, we recall the framework of the asymptotic quantization of general relativity at null infinity. In \cref{subsec:alg} we review the $\ast$-algebra quantization of asymptotic fields and BMS charges. In \cref{sec:mem-fock}, we review the construction of Fock representations with memory. Finally, \cref{sec:coherent-states} we illustrate how the (non-unique) ``memory vacua'' which appear in the construction of the memory Fock spaces can be viewed as ``generalized coherent states''. For a more detailed presentation, we refer the reader to \cite{narain81,asymp-quant,Ashtekar:2018lor,PSW-IR}. The reader familiar with these references may skip to \cref{sec:rigged}. 

\subsection{Asymptotic quantization algebra}
\label{subsec:alg}
The radiative degrees of freedom for the gravitational field are represented by the shear tensor $\sigma_{AB}(x)$ on $\scri$, where $x\in \scri$. Throughout this discussion, the symbol \(u\) signifies the retarded/advanced time coordinate on \(\scri\), and \(\hat x\) represents a point on \(\bb S^2\); we denote any point on \(\scri\) by $x=(u,\hat x)$. The uppercase Latin indices \(A,B,\ldots\) denote abstract indices for tensor fields on \(\bb S^2\). The Bondi News tensor is defined by
\be\label{eq:News-defn}
    N_{AB} \defn 2 \partial_u \sigma_{AB}
\ee
The asymptotic symplectic form on the radiative data at null infinity is given by (see \cite{AS-symp, GPS})
\be
    \Omega(\sigma_1,\sigma_2) & = \frac{1}{16\pi} \int_{\scri} d^{3}x~ \lb[ N_1^{AB} \sigma_{2AB} - N_2^{AB} \sigma_{1AB} \rb]
\ee
where \(d^3x \defn du d^2\hat x\) is the volume element on \(\scri\) with \(d^2\hat x\) the natural area element on \(\bb S^2\).

The asymptotic quantization algebra $\Alg$ is defined as the unital $\ast$-algebra generated by the identity $\op{1}$ together with the locally-smeared News operator
\be\label{eq:N(s)}
    \op{N}(s) \defn \int_{\scri}d^3x~ \op{N}_{AB}(x)s^{AB}(x),
\ee
where $s^{AB}$ is a real, smooth test tensor field of compact support (or rapid decay in \(u\)) on \(\scri\). The News operators satisfy the commutation relation
\be\label{eq:News-comm}
    [\op{N}(s_{1}), \op{N}(s_{2})] = -64\pi^{2}i \Omega(s_{1},s_{2})\op{1}
\ee
 
In the following, it will be convenient to instead use the Fourier space representation. For a test tensor, we define the Fourier transform by
\be
    s^{AB}(\omega, \hat p) = \frac{1}{\sqrt{2\pi}} \int_\scri d^3x~ e^{i \omega u} \delta_{\bb S^2}(\hat x, \hat p) s^{AB}(u,\hat x) 
\ee
where \(\delta_{\bb S^2}\) is the delta function on the \(2\)-sphere which only serves to relabel the point \(\hat x\) to the point \(\hat p\). We have made this relabelling for the following notational reason. We can identify the Fourier space coordinates \((\omega, \hat p)\) with a null vector in \(\bb R^4\) with a Minkowski metric \(\eta_{\mu\nu}\) such that in the standard Cartesian coordinates we have
\be\label{eq:p-defn}
    p^\mu = \omega (1,\hat p) \eqsp p^\mu p^\nu \eta_{\mu\nu} = 0
\ee
with \(\hat p \in \bb S^2\) being identified with a unit vector in the Euclidean \(\bb R^3\). Henceforth we will use the above identification and simply denote the Fourier space coordinates by \(p\).

The positive and negative frequency parts of \(s^{AB}(p)\) are defined by restricting it to \(\omega>0\) and \(\omega <0\), respectively. Note that, \(s^{AB}\) is a smooth test tensor with zero memory, its Fourier transform is smooth and finite at \(\omega = 0\), thus the decomposition into positive and negative frequency parts is well-defined. Further, since \(s^{AB}(u,\hat x)\) is real, we have that for \(\omega \geq 0\), \(s^{AB}(-\omega, \hat p) = \bar s^{AB}(\omega,\hat p)\). Thus, we can work exclusively with the positive frequency part of the test tensor. We denote this space of future-directed null vectors as \(C_+\) for the \emph{positive light cone}, and use the volume element \(d^3p = d\omega d^2\hat p\) on \(C_+\).

With these conventions, we can define the creation and annihilation operators directly in Fourier space by
\be
    \op a(s) \defn \int_{C_+}d^3p~ \op N_{AB}(p) \bar s^{AB}(p) \eqsp \op a(s)^* \defn \int_{C_+}d^3p~ \op{\bar N}_{AB}(p) s^{AB}(p)
\ee
Due to \cref{eq:News-comm}, these operators satisfy the commutation relations
\be\label{eq:ccr-smeared}
    \lb[ \op a(s_1) , \op a(s_2)^* \rb] = 16\pi \int_{C_+}d^3p~\omega~ \bar{s_1}^{AB}(p) s_{2AB}(p) \1
\ee
which in the ``unsmeared'' form is
\be\label{eq:ccr}
    \lb[ \op a_{AB}(p_1) , \op a_{CD}(p_2)^* \rb] = 16\pi \omega_1 \delta_{C_+}(p_1,p_2) \lb(q_{A(C} q_{D)B} - \half q_{AB} q_{CD}\rb) \1
\ee

States on the algebra $\Alg$ are linear maps $\Psi:\Alg \to \bb C$ such that $\Psi(\op{O}\op{O}^*)\geq 0$ for all $\op{O} \in \Alg$ and \(\Psi(\1) > 0 \). Furthermore, a state is normalized, if $\Psi(\op{1})=1$. Given that any element of $\Alg$ is a sum of products of the local news operator $\op{N}(s)$, specifying a state is equivalent to specifying its $n$-point correlation functions $\Psi(\op{N}(s_{1})\dots \op{N}(s_{n}))$. As explained in \cite{Kay:1988mu,Hollands:2014eia,PSW-IR} we also need to impose the \emph{Hadamard condition} on any state \(\Psi\) which in our notation takes the form\footnote{Since we will not always work with unit-normalized states there is a factor of \(\Psi(\1) > 0\) in the first term of \cref{eq:had-form-News}.}
\be \label{eq:had-form-News}
   \Psi(\op{N}_{AB}(x_{1})\op{N}_{CD}(x_{2}))= - 8 \frac{\lb(q_{A(C} q_{D)B} - \half q_{AB} q_{CD}\rb) \delta_{\mathbb{S}^{2}}(\hat x_{1},\hat x_{2})}{(u_{1}-u_{2}-i0^{+})^{2}} \Psi(\1) + S_{ABCD}(x_{1},x_{2})
\ee
where $S_{ABCD}$ is a (state-dependent) bi-tensor on $\scri^{-}$ that is symmetric in $A,B$ and in $C,D$, satisfies \(q^{AB}S_{ABCD} = q^{CD} S_{ABCD} = 0\) and is symmetric under the simultaneous interchange of $x_{1}$ with $x_{2}$ and the pair of indices $A,B$ with the pair $C,D$. Additionally we require that $S_{ABCD}$ as well as the connected $n$-point functions for $n\neq 2$ of a Hadamard state are sufficiently regular in the sense that they are smooth in retarded/advanced time, decay as  $O((\sum_{i}u_{i}^{2})^{-1/2-\epsilon})$ for some $\epsilon>0$ and are square-integrable in angles. These conditions are sufficient to ensure that the state has the same short distance singularity behavior as the vacuum and that the radiated energy at null infinity is finite. 

The News algebra \(\Alg\) described above can be extended to include the memory and the BMS charges which are the generators of infinitesimal BMS transformations on null infinity. We refer the reader to, e.g., \cite{PSW-IR} for further details of this construction, but recall the essential aspects. The classical gravitational memory is defined by
\be\label{eq:memory-defn}
    \Delta_{AB}(\hat x) \defn \frac{1}{2}\int_{\bb R} du~ N_{AB}(x) = \sigma_{AB}(u=\infty, \hat x) - \sigma_{AB}(u = -\infty, \hat x)
\ee
or in Fourier space by
\be\label{eq:memory-defn-fourier}
    \Delta_{AB}(\hat p) = \sqrt{\frac{2}{\pi}}\, N_{AB}(\omega=0, \hat p)
\ee
We will consider the smeared memory observable\footnote{The memory can be decomposed as $\Delta_{AB}=\big(\ms{D}_{A}\ms{D}_{B}-(1/2)q_{AB}\ms{D}^{2}\big)\alpha + \varepsilon_{(A}{}^{C}\ms{D}_{B)}\beta$ where $\alpha$ and $\beta$ are functions on $\bb{S}^{2}$ corresponding to the ``electric'' and ``magnetic'' parity parts respectively (see ,e.g., \S~III.D of \cite{Satishchandran_2019}). We only consider the electric parity part of the memory, though all of the constructions of this paper can be straightforwardly extended to the magnetic parity part.} 
\be
    \Delta(f) = \frac{1}{2}\int_{\bb R} d^3x~ N_{AB}(x) \ms D^A \ms D^B f(\hat x) = \sqrt{\frac{2}{\pi}}\, \int_{\bb S^2} d^2\hat p~ N_{AB}(0, \hat p) \ms D^A \ms D^B f(\hat p)
\ee
where \(f\) is a smooth function on \(\bb S^2\).

While it is tempting to define the memory operator in the quantum theory as the limit to zero frequency of the News operator, we must remember that the operators are essentially \emph{distributional}, and such a limit is not defined. Instead, we proceed as follows. We define the memory operator by the commutation relations one obtains from the classical Poisson brackets. The memory operator then commutes with the News operator and thus the creation and annihilation operators (see \cite{PSW-IR}). Now that the memory is, in fact, the zero frequency limit of the News is imposed as a condition on the states. For all Hadamard states the one-point function of the News is a smooth function in frequency. Since we can safely take the zero frequency value of a smooth function, we impose the condition that the one-point function of the memory operator for all allowed states is the zero frequency value of the one-point function of the News operator. Equivalently in position space we impose
\be\label{eq:mem-1pt-cond}
    \Psi(\op\Delta(f)) = \frac{1}{2} \int_{\scri} d^3x~ \Psi(\op N_{AB}(x)) \ms D^A \ms D^B f(\hat x)
\ee
as a condition on allowed states.\footnote{Since, by the Hadamard condition, $S_{ABCD}$ and the connected $n$-point functions for $n\neq 2$ are sufficiently regular that one could also impose analogous compatibility relations between the memory $n$-point functions and the smooth parts of the News $n$-point functions. However, for all states considered in \cref{sec:mem-fock}, these relations will be satisfied if the state satisfies \cref{eq:mem-1pt-cond}.}

The BMS charge observable that generates the infinitesimal BMS transformations is given by (see \cite{GPS})
\be\label{eq:bms-flux}
    {\mc Q}_{i^0} {(f,X)} = \frac{1}{16\pi} \int_{\scri} d^3x~ N^{AB} &\bigg[ \tfrac{1}{2} (f + \tfrac{1}{2}u \ms D_C X^C) N_{AB} + \ms D_A \ms D_B f + \Lie_X \sigma_{AB} - \tfrac{1}{2} (\ms D_C X^C) \sigma_{AB} \bigg].
\ee
where \(f\) is a smooth function on \(\bb S^2\) parametrizing the BMS supertranslations and \(X^A\) is a conformal Killing vector field on the \(2\)-sphere generating Lorentz transformations. For $X^A = 0$, the supertranslation charge, or supermomentum, ${\mc Q}_{i^0} {(f)}$, can be written as
\be\label{eq:QJ-mem}
    {\mc Q}_{i^0} (f) = {\mc J}(f) + \frac{1}{8\pi} {\Delta}(f)
\ee
where \(\Delta(f)\) is the memory observable and \be\label{eq:null-mem}
    {\mc J}(f) \defn \frac{1}{32\pi} \int_{\scri} d^3x~ f N^{AB} N_{AB} 
\ee
is called the {\em null memory}. The BMS charges can be upgraded to operators in the quantum theory (see \cite{PSW-IR}) and give the non-vanishing commutation relations
\be\label{eq:charge-comm}
    \lb[ \op{\mc Q}_{i^0}(f,X), \op N(s) \rb] &= i \op N(\delta_{(f,X)}s) \\
    [\op{\mc{Q}}_{i^{0}}(X_{1}),\op{\mc{Q}}_{i^{0}}(X_{2})] &= i \op{\mc{Q}}_{i^{0}}([X_{1},X_{2}]) \\
    [\op{\mc{Q}}_{i^{0}}(X),\op{\mc{Q}}_{i^{0}}(f)] &= i \op{\mc{Q}}_{i^{0}}\lb( \Lie_X f-\tfrac{1}{2}(\ms{D}_{A}X^{A})f \rb) \\
    [\op{\mc{Q}}_{i^{0}}(X),\op{\Delta}(f)] &= i \op{\Delta}\lb(\Lie_X f-\tfrac{1}{2}(\ms{D}_{A}X^{A})f\rb)
\ee
where
\be
    \delta_{(f,X)} s^{AB} & = \lb( f + \tfrac{1}{2} u \ms D_C X^C \rb) \partial_u s^{AB} + \Lie_X s^{AB} + \tfrac{1}{2} (\ms{D}_{C} X^{C}) s^{AB}
\ee
and \([X_1,X_2]\) denotes the Lie bracket of the two Lorentz vector fields on \(\bb S^2\).\\

Since one of the main goals of the paper is to find the eigenstates of the supermomentum charge it will be useful to have an expression for the BMS supermomentum in terms of the creation and annihilation operators. To do this begin with the BMS supermomentum charge and write it in terms of the memory and null memory as in \cref{eq:QJ-mem}
\be\label{eq:QJ-mem-op}
    \op{\mc Q}_{i^{0}}(f) = \op{\mc J}(f) + \frac{1}{8\pi} \op\Delta(f).
\ee
Now the memory has already been defined as an operator above, so all we need to do is to write the quantum expression corresponding to the null memory in terms of the News operator as in the classical expression \cref{eq:null-mem}. However, \cref{eq:null-mem} involves products of the News at coincident points which is, in general, not defined since quantum operators are distributional. Rigorously, the operators corresponding to such classical expressions can be defined using ``Hadamard regularization'' (see \cite{Hollands:2014eia}). However, by \cref{eq:had-form-News}, the short distance singular structure of any Hadamard state is universal and so this ``Hadamard regularization'' is equivalent to the usual ``normal ordering'' prescription of creation and annihilation operators in Fourier space. This gives the null memory operator in terms of the creation and annihilation operators as
\be\label{eq:null-mem-op}
    \op{\mc J}(f) \defn \frac{1}{16\pi} \int_{C^{+}}d^{3}p~f(\hat{p})\op{a}^{AB}(p)^{\ast}\op{a}_{AB}(p) 
\ee
and the commutation relation with the annihilation operator can be written as
\be\label{eq:J-a-comm}
    \lb[ \op{\mc J}(f) , \op a(p) \rb] = \omega f(\hat p) \op a(p)
\ee
where the equation is understood as smeared with \(\bar s^{AB}(p)\). Similarly, an expression for the Lorentz charges $\op{\mc{Q}}_{i^{0}}(X)$ can be given in terms of the creation and annihilation operators but we will not need it in what follows.

\subsection{Memory Fock space representations}
\label{sec:mem-fock}

We now recall the construction of the memory Fock spaces which are representations of the asymptotic algebra \(\Alg\), including the News, the memory and the supermomentum charge. The Lorentz charges cannot be defined as operators on these memory Fock spaces, except when the memory vanishes, but as shown in \cite{memory-orbits} one can construct Hilbert spaces on which the Lorentz charges are represented as operators by taking direct integrals over the memory Fock spaces. This construction is summarized at the end of this section.

We start with the BMS invariant vacuum state \(\s_0\), which is a Gaussian algebraic state with vanishing \(1\)-point function of the News and \(2\)-point function given  by \cref{eq:had-form-News} with \(S_{ABCD} = 0\) and the normalization \(\omega_0(\1) = 1\). Further \(\omega_0\) is also an eigenstate with vanishing eigenvalue of the memory and all the BMS charges. The corresponding Hilbert space, constructed by applying the creation operators $\op a(s)^*$ to the vacuum \(\ket{\s_0}\), is the Fock space $\Fock_{0}$:
\be\label{eq:fock-0}
    \Fock_{0}= \mathbb{C}\oplus \bigoplus_{n\geq 1}\underbrace{\big(\Hilb_{0}\otimes_{S}\dots \otimes_{S}\Hilb_{0}\big)}_{n\text{ times}}.
\ee
Here, $\otimes_{S}$ denotes the symmetrized tensor product, and the one-particle Hilbert space $\Hilb_{0}$ is the completion of the space of positive frequency test functions $s^{AB}$ with respect to the inner product:
\be\label{eq:innerprod}
    \braket{s_{1}|s_{2}}_{0} \defn \s_{0}(\op a(s_1) \op a(s_2)^*) = 16\pi \int_{C_+} d^3p~ \omega\, \bar s_{1}^{AB}(p) s_{2,AB}(p)\, .
\ee

It is well-known that the standard Fock space of gravitons provides a dense set of Hadamard states with finite energy, (super)momentum, and angular momentum. The action of the BMS charges on states in \(\Fock_0\) can be found by using their commutation relations (\cref{eq:charge-comm}) and noting that they annihilate the vacuum.\footnote{The Lorentz charge operators are well-defined on \(\Fock_0\) and can be used to define the helicity of the graviton states. The helicity of a \(1\)-graviton state in \(\Fock_0\) can be computed to be \(\pm 2\) as expected; see \cite{asymp-quant} for details.} However, the standard Fock space does not encompass all states relevant for scattering theory. Specifically, since the memory operator $\op{\Delta}$ annihilates $\ket{\s_{0}}$, the commutation of $\op{\Delta}$ with $\op N(s)$ implies that {\em all} states in $\Fock_{0}$ have zero memory.

As discussed in the introduction, the memory is generically not conserved in scattering and thus states in the standard zero-memory Fock space are not sufficient to formulate an infrared finite scattering theory. A large supply of states with non-zero memory can be constructed following the procedure in \cite{asymp-quant}. Choose a classical News tensor \(\nu_{AB}(p)\) with memory \(\Delta_{AB}(\hat p)\) and consider the map
\be\label{eq:aut-defn}
    \mf a_\nu [\op a(s)] = \op a(s) + \nu(\bar s) \1 \eqsp \mf a_\nu [\op \Delta(f)] = \op \Delta(f) + \Delta(\nu, f) \1
\ee
where
\be\label{eq:nu-smeared}
    \nu(\bar s) = \int_{C_+}d^3p~ \nu_{AB}(p) \bar s^{AB}(p)
\ee
and \(\Delta(\nu,f)\) denotes the chosen classical memory smeared with the function \(f\). This map can be extended to an automorphism \(\mf a_\nu : \Alg \to \Alg\) of the algebra containing the News and memory by demanding that it preserves the identity \(\1\), the product and the \(*\)-relation in the algebra. Then, define a new algebraic state \(\s_\nu\) by
\be\label{eq:aut-state}
    \s_\nu(\op O) = \s_0(\mf a_\nu[\op O]) \eqsp \text{for all } \op O \in \Alg
\ee
From \cref{eq:aut-defn,eq:aut-state}, it can be verified that the state \(\s_\nu\) is an eigenstate of the memory operator with eigenvalue \(\Delta(\nu;f)\). If the News tensor is chosen so that \(\Delta_{AB} = 0\), this automorphism can be represented by a unitary operator on \(\Fock_0\), and the new state \(\s_\nu\) is a coherent state in \(\Fock_0\) (see \cite{asymp-quant} and \cref{sec:coherent-states} below). However, when \(\Delta_{AB}\) is non-vanishing, this defines a new Gaussian state which does not lie in \(\Fock_0\) and is an eigenstate of the memory operator with eigenvalue equal to the chosen memory \(\Delta_{AB}\). 

Next we explain our regularity conditions on the News tensor \(\nu_{AB}\) used in \cref{eq:aut-defn,eq:aut-state}. Since we want the memory of the chosen News tensor to be finite, from \cref{eq:memory-defn}, it is sufficient to demand that \(\nu_{AB}\) is smooth in \(u\) and decays as \(O(1/u^{1+\epsilon})\) for large \(u\) and some \(\epsilon > 0\). Computing the \(2\)-point function of News operator in the state \(\s_\nu\), using \cref{eq:had-form-News,eq:aut-defn}, we see that this state satisfies the Hadamard condition \cref{eq:had-form-News} for the decay in \(u\). Further, using the automorphism \cref{eq:aut-defn} and \cref{eq:null-mem-op} we have
\be
    \s_\nu(\op{\mc J}(f)) = \frac{1}{16\pi} \int_{C_+}d^3p~ f(\hat p) \bar\nu^{AB}(p) \nu_{AB}(p) = \frac{1}{32\pi} \int_\scri d^3x~ f(\hat x) \nu^{AB}(x) \nu_{AB}(x)
\ee
where in the last equality we have written the expression in position space on \(\scri\). Note that for \(f(\hat x ) =1\) this is just the expected Bondi energy flux in the quantum state \(\s_\nu\). The integral in \(u\) is finite for the decay of the News we have already specified, and we also need that the \(\nu_{AB}(x)\) is square-integrable in \(\bb S^2\). Thus the corresponding memory \(\Delta_{AB}\) must also be square-integrable on the \(2\)-sphere. As mentioned above, and proved in \cref{sec:amplitudes} below, the condition on the memory cannot be strengthened to only include smooth memories since this is not generically preserved under scattering. Henceforth, we call the News tensor and corresponding state \(\s_\nu\) satisfying the above conditions as \emph{regular}.

Acting on the state \(\s_\nu\) with the smeared creation operators generates an entire new Fock space \(\Fock_\Delta\) which is \emph{unitarily inequivalent} to \(\Fock_0\). The choice of the ``vacuum'' state $\s_{\nu}$ in this construction is not unique --- if a different News tensor $\nu'_{AB}$ is picked with identical memory, i.e., $\Delta(\nu', f) = \Delta(\nu, f)$ for all \(f\), then our regularity conditions above ensure that the two representations will be unitarily equivalent. For this reason, the memory Fock spaces \(\Fock_\Delta\) are labeled by the memory instead of the chosen News tensor.

The memory of the radiative field is invariant under BMS supertranslations, i.e., for any chosen classical News \(\nu_{AB}\) the supertranslated News tensor \(\nu'_{AB}\) has the same memory. Thus, any BMS supertranslation maps a choice of vacuum \(\s_\nu\) to another equivalent one \(\s_{\nu'}\) within the same memory Fock space. Using the commutation relations with the creation operators, this action can be extended to the entire memory Fock space. Consequently, the BMS supermomentum is well-defined on any memory Fock space associated with a square-integrable memory tensor.

On the other hand, the Lorentz charges, in general, cannot be defined on a memory Fock space. Recall that the memory tensor is not Lorentz-invariant unless it vanishes, and the Lorentz charges do not commute with the memory operator. Since all the states in the memory Fock spaces are eigenstates of the memory operator, it follows that the Lorentz charges cannot be densely-defined operators on the Fock spaces of definite memory. Only a subgroup of the Lorentz transformations which preserve the memory tensor --- the `little group'' of the memory --- have well-defined charges on the given Fock space \(\Fock_\Delta\). However, since the memory is not preserved under scattering, the little groups of the in and out memory will not coincide in general. 

It was recently shown in \cite{memory-orbits} that plenty of states with finite Lorentz charges can be constructed by taking direct integrals over the memory Fock spaces. One could consider taking the direct integral over the infinite-dimensional space of all allowed memories. This scenario was studied in \cite{Herdegen_1997} and \S~7 of \cite{PSW-IR} and requires a choice of an infinite-dimensional measure. Though this choice isn't unique, \cite{PSW-IR} argued that there is no Gaussian measure over the memory space yielding a representation with well-defined energy and angular momentum. The alternative construction in \cite{memory-orbits} takes a direct integral over a finite-dimensional subspace of memories. This construction can summarized as follows. The space of memories obtained by applying the complete Lorentz group to a given memory $\Delta_{AB}$ forms a finite-dimensional manifold --- the \emph{orbit space} \(\orbit\) associated with the given memory \(\Delta_{AB}\). If \(y \in \orbit\) represents a point in the orbit space, the Lorentz transformation gives a set of Fock spaces $\Fock_{\Delta(y)}$ corresponding to each point in \(\orbit\). Each orbit space has a unique, Lorentz invariant measure \(\mu(y)\). Consequently, for each memory \(\Delta_{AB}\), we obtain a \emph{direct integral} Hilbert space
\be\label{eq:DIDelta}
    \Fock_{\textrm{DI},\Delta} = \int_{\orbit}^{\oplus}d\mu(y)\Fock_{\Delta(y)}
\ee
where the subscript ``\(\textrm{DI},\Delta\)'' indicates a direct integral. States in this direct integral Hilbert space have continuous memory distributions across the orbit space \(\orbit\). These Hilbert spaces admit a unitary representation of the complete asymptotic BMS symmetry group, and contain a dense set of states with non-zero memory, along with well-defined energy and angular momentum.

Thus, while there is no single, preferred Hilbert space which includes all the states that can arise in scattering, the memory Fock spaces and direct integrals obtained from them, provide plenty of physical states to use in an infrared finite scattering theory.

\subsection{Memory vacua as generalized coherent states}
\label{sec:coherent-states}

In the subsection, we illustrate that the memory vacua \(\s_\nu\) can be viewed as generalizations of coherent states and collect some formulae which will be useful in the main arguments of the paper.

From \cref{eq:aut-defn,eq:aut-state} it can be seen that the memory vacua constructed above are all eigenstates of the annihilation operator in the sense that \(\s_\nu(\op O \op a(s)) = \s_\nu(\op O) \nu(\bar s)\) and \(\s_\nu(\op a(s)^*\op O) = \s_\nu(\op O) \bar\nu(s)\) for all \(\op O \in \Alg\), where \(\nu(\bar s)\) is as given in \cref{eq:nu-smeared}. Thus, in any \(\Fock_\Delta\) we have
\be\label{eq:coherent-state-defn}
    \op a(s) \ket{\s_\nu} = \nu(\bar s) \ket{\s_\nu} \eqsp \norm{\s_\nu}^2_\Delta = 1
\ee
where the last relation follows from \cref{eq:aut-state} and that the automorphism \(\mf a_\nu\) preserves the identity operator. When the memory of \(\nu_{AB}\) vanishes \(\s_\nu\) is precisely a unit-normalized coherent state in \(\Fock_0\). For non-zero memory \(\s_\nu\) can be considered as a \emph{generalized coherent state} in \(\Fock_\Delta\). Two such coherent states in the same memory Fock space can be related by a unitary operator. For later, we will need the explicit form of this unitary, which we detail next.

Consider first \(\nu_{AB}(p) = -2i \omega s_{AB}(p)\) where \(s_{AB}(p)\) is a test tensor of compact support --- \(s_{AB}(p)\) is smooth at zero frequency and the memory of this radiation field vanishes. The corresponding normalized coherent state \(\ket{\s_\nu} \in \Fock_0\) is given by\footnote{Given \cref{eq:coherent-state-defn}, the unitary operator in \cref{eq:coherent-state-0} is only determined up to multiplication by a non-zero complex number; we have fixed this constant for convenience of the subsequent computations.}
\be\label{eq:coherent-state-0}
    \ket{\s_\nu} &= e^{-\tfrac{i}{8\pi} \op N(s)} \ket{\s_0} = e^{-\tfrac{1}{128\pi^2} \norm{s}^2_0} e^{-\tfrac{i}{8\pi} \op a(s)^*} \ket{\s_0}
\ee
where the second form follows from writing the News operator into creation and annihilation operators (the annihilation operators annihilate the vacuum \(\ket{\s_0}\)) and using the Baker-Campbell-Hausdorff formula, and similarly \cref{eq:coherent-state-defn} can be verified. This unitary operator represents the automorphism \(\mf a_\nu\) in \cref{eq:aut-defn} when \(\nu_{AB}\) has zero memory. Note that these unitary operators satisfy the \emph{Weyl relation}
\be\label{eq:weyl-relation}
    e^{-\tfrac{i}{8\pi} \op N(s')} e^{-\tfrac{i}{8\pi} \op N(s)} = e^{\half\, i\Omega(s',s)} e^{-\tfrac{i}{8\pi} \op N(s'+s)}
\ee
If \(\ket{\s_{\nu'}} \in \Fock_0\) is another coherent state with News \(\nu'_{AB}(p) = -2i \omega s'_{AB}(p)\) we have
\be\label{eq:coherent-change-0}
    \ket{\s_{\nu'}} 
    &= e^{-\tfrac{i}{8\pi} \op N(s')} e^{\tfrac{i}{8\pi} \op N(s)}  \ket{\s_\nu} \\
    &= e^{-\half\, i\Omega(s',s)}\, e^{-\tfrac{i}{8\pi} \op N(s'-s)}  \ket{\s_\nu} \\
    &= e^{-\half\, i\Omega(s',s) - \tfrac{i}{8\pi} \nu(\bar s'-\bar s) -\tfrac{1}{128\pi^2} \norm{s'-s}^2_0}\, e^{-\tfrac{i}{8\pi} \op a(s'-s)^*}\ket{\s_\nu}
\ee
and the inner product between any two coherent states in \(\Fock_0\) is
\be
    \braket{\s_\nu | \s_{\nu'}}_0 
    &= e^{-\half\, i\Omega(s',s) -\tfrac{i}{4\pi} \Re\lb(\nu(\bar s'-\bar s)\rb) -\tfrac{1}{128\pi^2} \norm{s'-s}^2_0}
\ee

We wish to generalize this to vacuum states \(\ket{\s_\nu}\) with memory. To do this, we note that for any regular News tensor \(\nu_{AB}\) with memory we can find a corresponding shear tensor \(\sigma_{AB}\) such that \(\nu_{AB}(p) = -2i \omega \sigma_{AB}(p)\). Note that this shear tensor \(\sigma_{AB}\) can be chosen to be smooth and decay rapidly at high frequencies but \emph{will not} be finite at zero frequency, due to non-vanishing memory. For this reason, we use the notation \(\sigma_{AB}\) to avoid confusion with the test tensors \(s_{AB}\) which are finite even at zero frequency. Next if we replace \(s_{AB}(p)\) and \(s'_{AB}(p)\) with shear functions \(\sigma_{AB}(p)\) and \(\sigma'_{AB}(p)\) that have the same memory \(\Delta_{AB} \neq 0\) then the last two lines of \cref{eq:coherent-change-0} are still finite since the symplectic form is well-defined for any shear functions with memory, and the difference \(\sigma'_{AB} - \sigma_{AB}\) has zero memory. Thus, for any given memory \(\Delta\) the change from one coherent state to another in \(\Fock_\Delta\) can be defined by the unitary
\be\label{eq:coherent-change}
    \ket{\s_{\nu'}} 
    &= e^{-\half\, i\Omega(\sigma',\sigma)} e^{-\tfrac{i}{8\pi} \op N(\sigma'-\sigma)}  \ket{\s_\nu} \\
    &= e^{-\half\, i\Omega(\sigma',\sigma)} e^{-\tfrac{i}{8\pi} \nu(\bar \sigma'-\bar \sigma)}  e^{-\tfrac{1}{128\pi^2} \norm{\sigma'-\sigma}^2_0} e^{-\tfrac{i}{8\pi} \op a(\sigma'-\sigma)^*}\ket{\s_\nu} 
\ee
and the inner product of two coherent states in \(\Fock_\Delta\) is given by
\be\label{eq:coherent-state-inp}
    \braket{\s_\nu | \s_{\nu'}}_\Delta &= e^{-\half\, i\Omega(\sigma',\sigma)} e^{-\tfrac{i}{4\pi} \Re \lb(\nu(\bar \sigma'-\bar \sigma)\rb)}  e^{-\tfrac{1}{128\pi^2} \norm{\sigma'-\sigma}^2_0}
\ee

\begin{remark}[Weyl algebra]
    It is possible to define an abstract Weyl algebra with operators of the form \(\op U(\sigma)\) where \(\sigma_{AB}\) denotes a shear tensor with memory. Equipped with the relations  \(\op U(\sigma)^* = \op U(-\sigma)\), \(\op U(0) = \1\) along with the analogue of \cref{eq:weyl-relation}
    \be\label{eq:weyl-relation-gen}
        \op U(\sigma') \op U(\sigma) = e^{\half\, i\Omega(\sigma',\sigma)} \op U(\sigma' + \sigma)
    \ee
    Note that together these imply that \(\op U(\sigma)^* = \op U(\sigma)^{-1}\), i.e. \(\op U(\sigma)\) is an abstract unitary operator. Further, when \(\sigma_{AB}\) has vanishing memory, they are equivalent to the ones defined in terms of the News operator above. But for a general \(\sigma_{AB}\) they cannot be interpreted as the exponential of any well-defined operator involving the News and must be defined abstractly. Further, for general \(\sigma_{AB}\), these generalized unitaries are not represented as operators on any memory Fock space \(\Fock_\Delta\) or even on direct integrals over memory orbits constructed in \cite{memory-orbits}. They can be represented on the direct sum over all memories of \(\Fock_\Delta\) which is a non-separable Hilbert space. This kind of strategy was used by Kibble \cite{Kibble1} --- along with a different convention and more formal rules for the exponential in \cref{eq:weyl-relation-gen} --- to analyze the problem of infrared divergences, but we will not need to introduce these abstract unitaries.
\end{remark}

\section{Improper eigenstates of supermomentum}
\label{sec:rigged}

In this section, we construct the eigenstates of the supermomenta in every memory Fock space $\Fock_{\Delta}$. The operator \(\op{\mc Q}_{i^0}(f)\) is a self-adjoint operator on $\Fock_{\Delta}$ and so the existence of such a complete set of eigenstates is guaranteed by the spectral theorem which yields a decomposition of all states in terms of projection-valued measures \cite{Reed-Simon}. However, for our purposes, we want to find simultaneous eigenstates of \(\op{\mc Q}_{i^0}(f)\) for all supertranslation functions \(f\). In terms of using the spectral theorem, this would require performing the daunting task of finding projection-valued measures on each \(\Fock_\Delta\) for all functions \(f\) on \(\bb S^2\). Thus purely for practical computational convenience, we will instead use the rigorous rigged Hilbert space formalism \cite{Gelfand4}.

As we will see, the construction of eigenstates of $\op{\mc{Q}}_{i^{0}}(f)$ on $\Fock_{\Delta}$ for $\Delta_{AB}\neq 0$, is far more non-trivial than the construction of eigenstates on $\Fock_{0}$. Indeed, in the standard Fock space $\Fock_{0}$, the supermomentum eigenstates can be straightforwardly constructed by noting that the vacuum $\ket{\omega_{0}}$ is a proper state annihilated by $\op{\mc Q}_{i^0}(f)$ for all $f$ (see remark \ref{rem:zero-memory-simplify}). Using the fact that $\ket{\s_{0}}$ has vanishing supermomentum and vanishing memory, \cref{eq:QJ-mem-op} together with \cref{eq:J-a-comm} imply that, formally, the $n$-particle momentum states 
\begin{equation}
\ket{p_{1}\dots p_{n}} \defn \frac{1}{\sqrt{n!}}\op{a}(p_{1})^* \cdots \op{a}(p_{n})^*\ket{\omega_{0}}
\end{equation}
are (improper) eigenstates of the supermomentum in $\Fock_{0}$ with eigenvalue 
\begin{equation}
\op{\mc{Q}}_{i^{0}}(f)\ket{p_1,\ldots,p_n} = \bigg(\sum_{i=0}^n \omega_i f(\hat{p}_{i}) \bigg)\ket{p_1,\ldots,p_n} \quad \quad \textrm{ (in $\Fock_{0}$)}
\end{equation}
However, this procedure cannot be directly generalized to construct eigenstates of $\op{\mc{Q}}_{i^{0}}(f)$ when the memory is non-zero. Indeed, as we will explicitly see, there is {\em no} state in $\Fock_{\Delta}$ that is annihilated by $\op{\mc{Q}}_{i^{0}}(f)$ and, furthermore, {\em all} of the eigenstates of $\op{\mc{Q}}_{i^{0}}(f)$ are improper states. The rigged Hilbert space formalism is particularly useful in this situation since it allows one to explicitly construct the (possibly improper) eigenstates of any self-adjoint operator.

In \cref{subsec:RiggedHilbertSpaces} we review the rigged Hilbert space formalism as well as the notion of improper eigenstates. In \cref{sec:bms-particles} we apply this formalism to to explicitly construct the eigenstates of the supermomentum and analyze their properties.

\subsection{ Rigged Hilbert spaces}
\label{subsec:RiggedHilbertSpaces}
As we have already stated, the eigenstates of the supermomentum are generically not proper states but will have a well-defined ``inner product'' with a dense set of states in the Hilbert space. Thus the eigenstates lie in the dual of this dense space. In this subsection, we will introduce the rigged Hilbert space, following \cite{Gelfand4}, which allows one to generally define these improper states as well as the notion of improper eigenstates. 

Consider a separable Hilbert space \(\Hilb\) and let \(\ms D\) be a dense vector subspace of \(\Hilb\). Furthermore, let \(\ms D\) also have the structure of a nuclear space (see \cite{Gelfand4,schaefer}). Let \(\ms D'\) be the dual of \(\ms D\) in the nuclear topology, i.e., the space of all continuous linear maps from \(\ms D\) to \(\bb C\). The dual of \(\Hilb\) in the Hilbert topology is isomorphic to \(\Hilb\) by the \emph{Riez-Frechet theorem} and so we have the inclusions
\be\label{eq:rigged-defn}
    \ms D \subset \Hilb \subset \ms D'
\ee
The triple of spaces \((\ms D, \Hilb, \ms D')\) is known as a \emph{rigged Hilbert space} or a \emph{Gelfand triple}. 

An \emph{improper state} \(F\) is then defined as an element of \(\ms D'\), i.e., 
\be
    F(\ket{\psi}) \in \bb C \text{ for all } \ket{\psi} \in \ms D
\ee
In the oft-used abuse of Dirac notation we will denote improper states also as \(\ket{F}\), and make the identification\footnote{The identification \cref{eq:dirac-not} between the linear map \(F\) and \(\ket{F}\) is antilinear so, \(\alpha F\) corresponds to \(\bar\alpha \ket{F}\) for any complex number \(\alpha\).}
\be\label{eq:dirac-not}
    F(\ket{\psi}) = \braket{F | \psi} \in \bb C \text{ for all } \ket{\psi} \in \ms D
\ee
that is, \(F(\ket{\psi})\) is viewed as the ``overlap'' of the improper state \(\ket{F}\) with the state \(\ket{\psi}\) in the dense subspace \(\ms D\). Note that in general \(\braket{F | \psi}\) is not defined if \(\ket{\psi}\) is a state in \(\Hilb\) but not in \(\ms D\). Similarly any expression of the form \(\braket{F_1 | F_2}\) need not be defined for improper states \(\ket{F_1}, \ket{F_2} \in \ms D'\).

Now, let \(\op O\) be some operator on \(\Hilb\). If \(\Hilb\) is infinite-dimensional then there need not exist any state in \(\Hilb\) which is an eigenstate of \(\op O\). However, one can find \emph{improper eigenstates} (or \emph{generalized eigenstates}) in the bigger space \(\ms D'\). Thus, we want to make sense of the expression
\be
    ``~ \op O \ket{F} = \lambda \ket{F} ~"
\ee
where \(\ket{F} \in \ms D'\) and \(\lambda \in \bb C\) is the eigenvalue. To do so, let the adjoint of \(\op O\) be such that \(\op O^\dagger\) maps the dense subspace \(\ms D\) to itself. An improper eigenstate \(\ket{F} \in \ms D'\) of \(\op O\) with eigenvalue \(\lambda\) is defined through\footnote{\Cref{eq:improper-eigenstate-defn} differs from the definition in \cite{Gelfand4}, since we wish to view the linear map \(F(\ket{\psi})\) as corresponding to \(\braket{F|\psi}\) according to \cref{eq:dirac-not}.}
\be\label{eq:improper-eigenstate-defn}
    F(\op O^\dagger\ket{\psi}) = \braket{ F | \op O^\dagger | \psi} = \bar\lambda \braket{F | \psi}  \eqsp \text{ for all } \ket{\psi} \in \ms D
\ee
We note that if \(\ket{F}\) is an improper eigenstate of \(\op O\) in the above sense then, so is \(\alpha \ket{F}\) for any complex number \(\alpha\), and their eigenvalues are the same. If the eigenstate \(\ket{F}\) happens to lie in \(\Hilb\), then we can fix this constant by fixing the normalization of the state. Similarly, if the Hilbert space contains a preferred state \(\ket{\psi}\) in \(\ms D\) then one can fix the constant ambiguity by prescribing the value of \(\braket{F|\psi}\), as long as this does not vanish. But in general, there is always the ambiguity of a constant factor in defining improper eigenstates.

An example of the rigged Hilbert space construction, with relevance to quantum mechanics, is where \(\Hilb \cong L^2(\bb R)\) is the Hilbert space of square-integrable functions on the real line, the dense subspace \(\ms D\) is the space of smooth compactly supported (or rapid decay at \(\infty\)) and \(\ms D'\) is the space of distributions (or tempered distributions, respectively). If \(\ket{\psi} \in \ms D\) corresponds to the wave function \(\psi(x)\), the improper position and momentum states \(\ket{x_0}\) and \(\ket{p_0}\) are then defined by
\be
    \braket{x_0| \psi} \defn \int_{\bb R}dx~ \delta(x-x_0) \psi(x) = \psi(x_0) \eqsp \braket{p_0 | \psi} \defn \frac{1}{\sqrt{2\pi}} \int_{\bb R} dx~ e^{ip_0x} \psi(x) = \psi(p_0)
\ee
where \(\psi(p)\) is the Fourier transform of \(\psi(x)\). These improper states \(\ket{x_0}\) and \(\ket{p_0}\) are then the improper eigenstates of the position and momentum operators, respectively.
\subsection{BMS particles as improper states on memory Fock spaces}
\label{sec:bms-particles}

In this section, we give the explicit construction of the BMS particle states as improper eigenstates of the supermomenta \(\op{\mc Q}_{i^0}(f)\) in the rigged Hilbert space setting. We will obtain a collection of improper states for any chosen memory
\be
    \ket{\ground;\Delta} \eqsp \ket{p_1,\ldots,p_n;\Delta} \eqsp \text{ for } p_i \in C_+
\ee
When the memory \(\Delta\) is zero, these coincide with \(\ket{\ground;0} = \ket{\s_0}\) being the BMS-invariant vacuum and \(\ket{p_1,\ldots,p_n; 0}\) are the usual improper momentum basis in \(\Fock_0\). Note the only proper state in this collection is \(\ket{\s_0}\), and all other states are only defined in the sense of improper states explained above. The main results of this section are collected in theorems \ref{thm:eigenstates} and \ref{thm:completeness} as well as lemma \ref{lem:soft-factor} below. \\

We start with the Fock space \(\Fock_\Delta\) with memory \(\Delta_{AB}\) and pick \(\ket{\s_\nu}\) to be some choice of a \emph{normalized} vacuum in \(\Fock_\Delta\) corresponding to a regular News tensor \(\nu_{AB}(p)\) (see \cref{sec:mem-fock,sec:coherent-states}). The construction below also holds for \(\Delta = 0\) with \(\ket{\s_\nu}\) being \emph{any} coherent state in \(\Fock_0\) --- not necessarily the invariant vacuum \(\ket{\s_0}\).

Within \(\Fock_\Delta\) consider the states \(\ket{\psi}\) (with finite norm) of the form 
\be\label{eq:nice-state}
    \ket{\psi} = \sum_{m=0}^\infty \frac{1}{\sqrt{m!}}\int_{C_+^m} d^3p_1 \ldots d^3p_m~ \psi^{A_1B_1\ldots A_mB_m} (p_1,\ldots,p_m) \op a_{A_1B_1}(p_1)^* \cdots \op a_{A_mB_m}(p_m)^* \ket{\s_\nu}
\ee
where \(\psi \in \bb C\) for \(m=0\), and each \(\psi^{A_1B_1\ldots A_mB_m} (p_1,\ldots,p_m)\) is a smooth test tensor on \(C_+^m\) of compact support (or rapid decay at large frequencies \(\omega_i\)). The linear span of such states forms a dense subspace \(\ms D_\Delta \subset \Fock_\Delta\). This dense space comes equipped with a nuclear topology induced from the topology on the space of smooth test tensors of compact support (see, e.g. \cite{schaefer}). For any fixed state \(\ket{\psi} \in \ms D_\Delta\), the right-hand side of \cref{eq:nice-state} can be rewritten in the same form relative to any other choice of regular vacuum \(\ket{\s_{\nu'}}\) using \cref{eq:coherent-change}. Thus, while we have chosen a vacuum to write the expression \cref{eq:nice-state}, the dense subspace \(\ms D_\Delta\) and its topology is independent of the choice of vacuum. Further, all the regular coherent vacuum states lie in \(\ms D_\Delta\).

Now we wish to find improper eigenstates of \(\op{\mc Q}_{i^0}(f)\) in the dual space \(\ms D'_\Delta\). So let \(\ket{F} \in \ms D'_\Delta\) be such an eigenstate. From \cref{eq:QJ-mem-op} we have
\be\label{eq:QJ-mem-eigenstate}
    \braket{F | \op{\mc Q}_{i^0}(f) | \psi} = \braket{F | \op{\mc J}(f) | \psi} + \Delta(f) \braket{F | \psi} \text{ for all } \ket{\psi} \in \ms D_\Delta
\ee
where we have used the fact that \(\op\Delta(f)\ket{\psi} = \Delta(f) \ket{\psi}\) for all \(\ket{\psi} \in \Fock_\Delta\). Thus, the eigenstates of \(\op{\mc Q}_{i^0}(f)\) are the same as the eigenstates of the null memory \(\op{\mc J}(f)\), and the eigenvalues differ only by the addition of \(\Delta(f)\) which is a constant on \(\Fock_\Delta\). Next, we construct these improper eigenstates of \(\op{\mc J}(f)\) and study some of their interesting properties. We will first define these states directly by their action on states in the dense subspace \(\ms D_\Delta\) as linear maps, and then show that these are indeed improper eigenstates of \(\op{\mc J}(f)\).

To include states with non-smooth but square-integrable memories, we will need to generalize the notion of improper states. Since the memory is not necessarily smooth, but only square-integrable, the quantities \(\braket{p_1,\ldots,p_n; \Delta | \psi}\) will not have a value at every point in the set \((p_1,\ldots,p_n) \in C_+^n\). However, this ``overlap'' will be a function on \(C_+^n\) which is smooth in frequencies and a square-integrable in the angular directions of the momenta. 

First define the ``BMS zero-particle'' improper state \(\ket{\ground;\Delta} \in \ms D'_\Delta\) by the following equations
\be\label{eq:0-plane-wave-defn}
    \braket{\ground;\Delta | \s_\nu} \defn A_\nu \eqsp \braket{\ground;\Delta | \op a_{A_1B_1}(p_1)^* \cdots \op a_{A_mB_m}(p_m)^* | \s_\nu} \defn 0 ~\text{ for all } m > 0
\ee
where \(A_\nu \neq 0 \in \bb C\) is a constant. Now define the ``BMS \(n\)-particle'' states \(\ket{p_1,\ldots,p_n;\Delta}\), for \(n > 0\), by
\be\label{eq:n-plane-wave-defn}
    \braket{p_1,\ldots,p_n;\Delta |\psi} \defn \frac{1}{\sqrt{n!}} \braket{\ground;\Delta | \op a_{A_1B_1}(p_1) \cdots \op a_{A_nB_n}(p_n) | \psi} \eqsp \text{ for all } \ket{\psi} \in \ms D_\Delta
\ee
which is the rigorous version of the relation
\be
    ``~ \ket{p_1,\ldots,p_n;\Delta} = \frac{1}{\sqrt{n!}} \op a_{A_1B_1}(p_1)^* \cdots \op a_{A_nB_n}(p_n)^* \ket{\ground;\Delta}  ~"
\ee
For notational convenience, we have suppressed the tensor indices on the left-hand side of \cref{eq:n-plane-wave-defn} above. In appendix \ref{sec:approx-eigenstates}, we prove that the BMS $0$-particle state $\ket{\ground;\Delta}$ can be approximated by a sequence of coherent states $\Psi_{\epsilon}$ in $\Fock_{\Delta}$ as $\epsilon\to 0$ and, in this limit, corresponds to an (improper) coherent state supported entirely at zero frequency with memory $\Delta_{AB}$. Recall that improper eigenstates are ambiguous up to multiplication by a constant. In \cref{eq:n-plane-wave-defn} we have fixed the constant ambiguity in \(\ket{p_1,\ldots,p_n;\Delta}\) for any $n\geq 1$ by defining them relative to \(\ket{\ground;\Delta}\). However, the constant \(A_\nu \neq 0\) which defines \(\ket{\ground;\Delta}\) cannot be fixed in general. We will frequently consider \(\ket{\ground;\Delta}\) as the \(n=0\) case of \(\ket{p_1,\ldots,p_n;\Delta}\). 

Next, we show that the BMS \(n\)-particle states defined in \cref{eq:n-plane-wave-defn} are well-defined improper states in the sense described above.
\begin{prop}
    The BMS \(n\)-particle states \(\ket{p_1,\ldots,p_n;\Delta}\) (for \(n \geq 0\)) are well-defined improper states in the sense that for all \(\ket{\psi} \in \ms D_\Delta\)
    \begin{enumerate}
    \item for \(n=0\), \(\braket{\ground;\Delta | \psi} \in \bb C\), thus \(\ket{\ground;\Delta} \in \ms D'_\Delta\).
    \item for \(n >0\), the family \(\braket{p_1,\ldots,p_n;\Delta | \psi} \) for \((p_1,\ldots,p_n) \in C_+^n\) defines a regular complex-valued test tensor on \(C_+^n\) which is square-integrable along the angular directions \(\hat p_1,\ldots,\hat p_n\) of the null momenta.
    \end{enumerate}
\begin{proof}
It follows directly from \cref{eq:0-plane-wave-defn} that for all \(\ket{\psi} \in \ms D_\Delta\) we have
\be\label{eq:p0-nice-state}
     \braket{\ground;\Delta | \psi} = A_\nu \psi \eqsp \braket{\ground;\Delta | \op a_{A_1B_1}(p_1)^* \cdots \op a_{A_nB_n}(p_n)^* | \psi} = 0
\ee
where on the right-hand side of the first equation above,  \(\psi \in \bb C\) is the \(m=0\) coefficient in \cref{eq:nice-state}. Thus, \(\ket{\ground;\Delta} \in \ms D'_\Delta\).

    From \cref{eq:coherent-state-defn,eq:n-plane-wave-defn} we have
    \be
        \braket{p_1,\ldots,p_n;\Delta | \s_\nu} &= \frac{1}{\sqrt{n!}}  \braket{\ground;\Delta | \op a_{A_1B_1}(p_1) \cdots \op a_{A_nB_n}(p_n) | \s_\nu} = \frac{A_\nu}{\sqrt{n!}} \nu_{A_1B_1}(p_1) \cdots \nu_{A_nB_n}(p_n)
    \ee
    Next, for any \(\ket{\psi} \in \ms D_\Delta\) we have
    \be
        \braket{p_1,\ldots,p_n;\Delta | \psi} &= \sum_{m=0}^\infty \frac{1}{\sqrt{n! m!}} \int_{C_+^m} d^3p'_1 \ldots d^3p'_m~ \psi^{A'_1B'_1\ldots A'_mB'_m}(p'_1,\ldots, p'_m) \\
        &\qquad \times \braket{\ground;\Delta | \op a_{A_1B_1}(p_1) \cdots \op a_{A_nB_n}(p_n) \op a_{A'_1B'_1}(p'_1)^* \cdots \op a_{A'_mB'_m}(p'_m)^* | \s_\nu}
    \ee
    Now we use the commutation relations \cref{eq:ccr} (or equivalently Wick's theorem) to put all the annihilation operators to the right of the creation ones. When \(n < m\) one is always left with an expression where all creation operators act on \(\bra{\ground;\Delta}\) from the right, which vanishes (from \cref{eq:p0-nice-state}); the remainder of the terms gives
\be\label{eq:pn-nice-state}
    \braket{p_1,\ldots,p_n;\Delta | \psi} 
    & = A_\nu \sum_{m=0}^n (16\pi)^m \sqrt{\frac{n!}{m!}} \frac{1}{(n-m)!} \Sym \bigg[\omega_1 \cdots \omega_m  \psi_{A_1B_1\ldots A_mB_m}(p_1,\ldots,p_m) \\
    &\qquad\qquad\qquad \times \nu_{A_{m+1}B_{m+1}}(p_{m+1}) \cdots \nu_{A_nB_n}(p_n) \bigg]
\ee
where the notation \(\Sym[\cdots]\) means total symmetrization of the expression in the brackets in the momenta \(p_1,\ldots,p_n\) and the corresponding tensor indices. For example, for \(n=2\)
\be\label{eq:sym-eg}
    \Sym \lb[\omega_1 \psi_{A_1B_1}(p_1) \nu_{A_2B_2}(p_2) \rb] = \frac{1}{2!} \lb[\omega_1 \psi_{A_1B_1}(p_1) \nu_{A_2B_2}(p_2) + \omega_2 \psi_{A_2B_2}(p_2) \nu_{A_1B_1}(p_1) \rb]
\ee
Since the News tensor \(\nu_{AB}\) is regular, in the sense described above, \cref{eq:pn-nice-state} defines a regular complex-valued tensor field on \(C_+^n\) which is square-integrable along the angular directions of the null momenta. When the memory is smooth --- and not merely in \(L^2(\bb S^2)\) --- then the News tensor can also be chosen to be smooth, and value of \(\braket{p_1,\ldots,p_n; \Delta | \psi}\) is defined at every \((p_1,\ldots,p_n) \in C_+^n\).
\end{proof}
\end{prop}
\Cref{eq:p0-nice-state,eq:pn-nice-state} give the explicit expression for the ``overlap'' of the BMS particles with proper states in \(\ms D_\Delta\). 

\begin{remark}[Change of vacuum]\label{rem:change-vacuum}
The only place that the choice of vacuum enters is in the constant \(A_\nu = \braket{\ground;\Delta | \s_\nu} \neq 0\) which, as we have argued before, cannot be fixed as there is no preferred choice of vacuum in \(\Fock_\Delta\). If \(\ket{\s_{\nu'}}\) is another choice of vacuum in \(\Fock_\Delta\) then we have from \cref{eq:coherent-change}
\be\label{eq:coherent-constant-change}
    A_{\nu'} = e^{-\half\, i\Omega(\sigma',\sigma) -\tfrac{i}{8\pi} \nu(\bar \sigma'-\bar \sigma) -\tfrac{1}{128\pi^2} \norm{\sigma'-\sigma}^2_0} A_\nu \neq 0
\ee
and for all \(n > 0\),
\be
    \braket{\ground;\Delta | \op a_{A_1B_1}(p_1)^* \cdots \op a_{A_nB_n}(p_n)^* | \s_{\nu'}} 
    & \propto \braket{\ground;\Delta | \op a_{A_1B_1}(p_1)^* \cdots \op a_{A_nB_n}(p_n)^* e^{i \op a(s'-s)^*} | \s_\nu} \\
    & = 0
\ee 
Similarly, for any fixed state \(\ket{\psi} \in \ms D_\Delta\), the right-hand side of \cref{eq:nice-state} can be rewritten in the same form relative to any other choice of regular vacuum \(\ket{\s_{\nu'}}\) using \cref{eq:coherent-change}. And it can be checked, though the computation is tedious, that for any choice of vacuum, we get the same \(\braket{p_1,\ldots,p_n; \Delta | \psi}\) for a given \(\ket{\psi}\).
\end{remark}

\begin{remark}[Zero memory case]
\label{rem:zero-memory-simplify}
    The expressions for the BMS particle states simplify considerably in the case the memory vanishes (\(\Delta = 0\)) and we choose the invariant vacuum \(\ket{\s_\nu} = \ket{\s_0}\). In this case, \cref{eq:nice-state} reduces to the Fock decomposition of the state in terms of \(n\)-particle sectors (\cref{eq:fock-0}). The definition \cref{eq:0-plane-wave-defn} implies that \(\ket{\ground;0} = \ket{\s_0}\) which is a proper state in this case. This allows us to normalize the improper state choosing \(A_0 = 1\). Then, \cref{eq:n-plane-wave-defn} implies that \(\ket{p_1,\ldots,p_n;0} = \ket{p_1,\ldots,p_n}\) are the usual \(n\)-particle momentum improper basis states, and \cref{eq:pn-nice-state} reduces to only one term
    \be
        \braket{p_1,\ldots,p_n; 0 | \psi} = (16\pi)^m \delta_{n,m} ~ \omega_1 \cdots \omega_m  \psi_{A_1B_1\ldots A_mB_m}(p_1,\ldots,p_m)
    \ee
Note that the above expression corresponds directly to the Fourier space version of the \(n\)-particle wave functions of \(\ket{\psi}\). In this sense, the \(\ket{p_1,\ldots,p_n; 0}\) can be interpreted as ``\(n\)-particle plane waves''. However, this interpretation cannot be made when the memory is non-zero --- see appendix \ref{sec:approx-eigenstates} for the interpretation of these improper states.
\end{remark}

In the following theorem, we show that the BMS \(n\)-particle states defined above are indeed improper eigenstates of the BMS supermomentum (and the null memory operator) in the sense of \cref{eq:improper-eigenstate-defn}.
\begin{thm}[Improper eigenstates of supermomentum]\label{thm:eigenstates}
    The improper states \(\ket{\ground;\Delta}\) and \(\ket{p_1,\ldots,p_n;\Delta}\) defined by \cref{eq:0-plane-wave-defn,eq:n-plane-wave-defn} are improper eigenstates of the null memory \(\op{\mc J}(f)\). The eigenvalue of \(\ket{\ground;\Delta}\) is zero, while that of \(\ket{p_1,\ldots,p_n;\Delta}\) is \(\sum_{i=1}^n \omega_i f(\hat p_i) \). From \cref{eq:QJ-mem-eigenstate}, it then follows that these are also improper eigenstates of the supermomentum \(\op{\mc Q}_{i^0}(f)\) with the eigenvalues shifted by \(\Delta(f)\).
\begin{proof}
Consider the null memory operator \cref{eq:null-mem-op} acting on a choice of regular vacuum \(\ket{\s_\nu} \in \ms D_\Delta\)
\be\label{eq:J-coherent}
    \op{\mc J}(f) \ket{\s_\nu} = \frac{1}{16\pi} \int_{C_+}d^3p~ f(\hat p) \nu^{AB}(p) \op a_{AB}(p)^* \ket{\s_\nu} = \frac{1}{16\pi} \op a(f\nu)^* \ket{\s_\nu}
\ee
where we note that \(f(\hat p) \nu_{AB}(p)\) can be used as a test tensor since it is finite at zero frequency, and the smeared creation operator above acting on \(\ket{\s_\nu}\) gives a proper state in \(\ms D_\Delta\). For \(m \geq 0\), we compute
\be
    &\braket{\ground;\Delta | \op{\mc J}(f) \op a_{A_1B_1}(p_1)^* \cdots \op a_{A_mB_m}(p_m)^* | \s_\nu} \\
     &\qquad\qquad = \frac{1}{16\pi} \braket{\ground;\Delta | \op a_{A_1B_1}(p_1)^* \cdots \op a_{A_mB_m}(p_m)^* \op a(f \nu)^* | \s_\nu} \\
     &\qquad\qquad\qquad + \lb( \sum_{i=1}^m \omega_i f(\hat p_i) \rb)  \braket{\ground;\Delta | \op a_{A_1B_1}(p_1)^* \cdots \op a_{A_mB_m}(p_m)^* | \s_\nu} \\
     &\qquad\qquad = 0 \,.
\ee
where we have repeatedly used the commutation relation \cref{eq:J-a-comm}. Here we note that for \(m=0\) the last term in the first equality above does not exist. Thus, \(\braket{\ground;\Delta | \op{\mc J}(f) | \psi} = 0\) for all supertranslations \(f\) and all \(\ket{\psi} \in \ms D_\Delta\), and \(\ket{\ground;\Delta}\) is an improper eigenstate of \(\op{\mc J}(f)\) with eigenvalue zero. 

Now let \(\ket{\psi} \in \ms D_\Delta\) and compute for \(n  > 0\)
\be
    \braket{p_1,\ldots,p_n;\Delta | \op{\mc J}(f) | \psi} & = \tfrac{1}{\sqrt{n!}} \braket{\ground;\Delta | \op a_{A_1B_1}(p_1) \cdots \op a_{A_nB_n}(p_n) \op{\mc J}(f) | \psi} \\
    & = \tfrac{1}{\sqrt{n!}}  \braket{\ground;\Delta | \op{\mc J}(f) \op a_{A_1B_1}(p_1) \cdots \op a_{A_nB_n}(p_n) | \psi} \\
    &\quad + \lb( \sum_{i=1}^n \omega_i f(\hat p_i) \rb) \tfrac{1}{\sqrt{n!}} \braket{\ground;\Delta | \op a_{A_1B_1}(p_1) \cdots \op a_{A_nB_n}(p_n) | \psi} \\
    & = \lb( \sum_{i=1}^n \omega_i f(\hat p_i) \rb) \braket{p_1,\ldots,p_n;\Delta | \psi}
\ee
where in the second line we have used the commutator \cref{eq:J-a-comm} and the last line uses that \(\ket{\ground;\Delta}\) is an eigenstate with zero eigenvalue. This shows that \(\ket{p_1,\ldots,p_n;\Delta}\) is an improper eigenstate of eigenvalue \(\sum_{i=1}^n \omega_i f(\hat p_i)\) as claimed.
\end{proof}
\end{thm}

Similar to the zero memory Fock space, the BMS particles form a complete set of improper basis for the Fock space \(\Fock_\Delta\), as shown in the following.
\begin{thm}[Completeness]
\label{thm:completeness}
    If \(\braket{p_1,\ldots,p_n;\Delta | \psi} = 0 \) for all \((p_1,\ldots,p_n) \in C_+^n\) and all \(n \geq 0\) for some \(\ket{\psi} \in \ms D_\Delta\) then, \(\ket{\psi} = 0\). Since the space \(\ms D_\Delta\) is dense in \(\Fock_\Delta\), the BMS particles form a complete improper basis for \(\Fock_\Delta\).
\begin{proof}
    First note that \(\braket{\ground;\Delta | \psi} = 0\) implies \(\psi = 0\). Then, from \cref{eq:pn-nice-state} we see that \(\braket{p_1,\ldots,p_n | \psi }\) depends only on the test tensors \(\psi^{A_1B_1\ldots A_mB_m}(p_1,\ldots,p_m)\) for \(m \leq n\). It follows from a straightforward inductive argument that \(\braket{p_1,\ldots,p_n | \psi } = 0\) for all \(0 \leq n \leq N\) implies that \(\psi^{A_1B_1\ldots A_mB_m}(p_1,\ldots,p_m) = 0\) for all \(0 \leq m \leq N\).
    
    Using another inductive argument in \(N\), it follows that \(\braket{p_1,\ldots,p_n;\Delta | \psi} = 0 \) for all \((p_1,\ldots,p_n) \in C_+^n\) and all \(n \geq 0\), implies that \(\psi^{A_1B_1\ldots A_mB_m}(p_1,\ldots,p_m) = 0\) for all \(m \geq 0\) and thus \(\ket{\psi} = 0\) in \(\ms D_\Delta\).
\end{proof}
\end{thm}

In the following lemma we show that the action of the improper BMS \(n\)-particle state \(\ket{p_1,\ldots,p_n; \Delta}\) on any proper state \(\ket{\psi}\) ``factorizes'' into the action of a BMS \((n-1)\)-particle state as we send the frequency of one of the momenta (say \(\omega_n\)) to zero. This ``soft factorization of the BMS particles'' will play a crucial role in showing that the analog of the soft factorization given by the Weinberg soft theorem holds in infrared finite scattering amplitudes in theorem~\ref{thm:soft}.

\begin{lemma}[Soft factorization]
For any BMS particle state \(\ket{p_1,\ldots,p_n}\) with \(n>0\) and any state \(\ket{\psi} \in \ms D_\Delta\) we have
\label{lem:soft-factor}
    \be
    \lim_{\omega_n \to 0} \braket{p_1,\ldots,p_n;\Delta | \psi} = \sqrt{\frac{2}{\pi n}}~ \Delta_{A_nB_n}(\hat p_n) \braket{p_1,\ldots,p_{n-1};\Delta | \psi}
    \ee
    When \(n=1\) the BMS particle state on the right-hand side is \(\bra{\ground;\Delta}\).
\begin{proof}
    We start with \cref{eq:pn-nice-state} and take the limit \(\omega_n \to 0\). Since each \(\psi^{A_1B_1\ldots A_mB_m}\) is a test tensor, the only terms left in the limit are those where \(p_n\) appears as an argument of the regular News tensor \(\nu_{A_nB_n}\). Since \(\nu_{AB}(p)\vert_{\omega = 0} = \sqrt{2/\pi} \Delta_{AB}(\hat p)\), taking into account the combinatorial factors, we are left with
\be
    &\lim_{\omega_n \to 0}\braket{p_1,\ldots,p_n;\Delta | \psi} \\
    &\qquad = \sqrt{\frac{2}{\pi n}}~ \Delta_{A_nB_n}(\hat p_n) A_\nu \sum_{m=0}^{n-1} (16\pi)^m \sqrt{\tfrac{(n-1)!}{m!}} \tfrac{1}{(n-1-m)!} \Sym \bigg[\omega_1 \cdots \omega_m  \psi_{A_1B_1\ldots A_mB_m}(p_1,\ldots,p_m) \\
    &\qquad\qquad\qquad\qquad \times  \nu_{A_{m+1}B_{m+1}}(p_{m+1}) \cdots \nu_{A_{n-1}B_{n-1}}(p_{n-1}) \bigg] \\
    &\qquad = \sqrt{\frac{2}{\pi n}}~ \Delta_{A_nB_n}(\hat p_n) \braket{p_1,\ldots,p_{n-1};\Delta | \psi}
\ee
as claimed.
\end{proof}
\end{lemma}
Note that for vanishing memory \(\Delta_{AB} = 0\), the above lemma is trivial. This is expected, since none of the states in \(\Fock_0\) have memory and so do not have any non-trivial behavior at zero frequency. Further, taking multiple limits as the frequencies go to zero, it is straightforward to show the following:
\begin{corollary}
\label{cor:infinite-bms-particles}
    If \(\Delta \neq 0\), and \(\braket{p_1,\ldots,p_N;\Delta | \psi} = 0\) for some \(N>0\) and \(\ket{\psi} \in \ms D_\Delta\) then, \(\braket{p_1,\ldots,p_n;\Delta | \psi} = 0\) for all \(n < N\).
\end{corollary}
The above corollary can be interpreted as saying that every state in the dense subspace \(\ms D_\Delta\) with non-zero memory is necessarily supported on infinitely many number of BMS particles. For suppose that \(\ket{\psi} \in \ms D_\Delta\) is supported only on finitely many BMS particles, i.e., there exists a \(N\) such that \(\braket{p_1,\ldots,p_n;\Delta | \psi} = 0\) for all \(n \geq N\). Then, from corollary \ref{cor:infinite-bms-particles}, we must have \(\braket{p_1,\ldots,p_n;\Delta | \psi} = 0\) for all \(n\), and consequently \(\ket{\psi} = 0\) by completeness of the BMS particle basis (theorem \ref{thm:completeness}).

We note here that states with non-zero memory and a well-defined action of all the BMS charge operators obtained in \cite{memory-orbits} are constructed by forming direct integrals over the definite memory Fock spaces \(\Fock_\Delta\). Since the BMS particles form a complete improper basis for any \(\Fock_\Delta\), they also form a complete basis for states constructed in \cite{memory-orbits}.\\

\begin{remark}[Relation to irreducible representations of the BMS group]
\label{rem:bms-particles-mccarthy}
The irreducible unitary representations of the BMS group were analyzed by McCarthy \cite{McCarthy1,McCarthy2,McCarthy-nucl}, by generalizing Wigner's method of induced representations \cite{Wigner}. McCarthy's construction can be summarized as follows: Given a supermomentum function \(\mc Q_{i^0}(\hat p)\),\footnote{For massive fields we can take the supermomentum to be square-integrable on the sphere \cite{McCarthy1,McCarthy2}. In this case, it is known that McCarthy's construction yields all unitary irreducible representations of the BMS group \cite{Piard}. Since we are dealing with massless fields, as also follows from \cref{thm:eigenstates}, we must consider the case where the supermomenta are distributions on the \(2\)-sphere \cite{McCarthy-nucl}. In this case, not all non-compact infinite discrete subgroups are known and thus McCarthy's classification is not complete. Similarly, it is not known if McCarthy's induced representations are all possible irreducible representations.}, let \(\grp{L}_{\mc Q_{i^0}}\) be the \emph{little group} of \(\mc Q_{i^0}\) --- i.e., the maximal subgroup of the Lorentz group which keeps the chosen supermomentum invariant. The {\em orbit space} \(\orbit_{\mc Q_{i^0}}\cong \grp{SL}(2,\bb{C})/\grp{L}_{\mc Q_{i^0}}\) parameterizes the space of supermomenta obtained by acting with the Lorentz group on the chosen \(\mc Q_{i^0}\) and admits a unique, Lorentz invariant measure. At each point in $\ms{O}_{\mc{Q}_{i^{0}}}$, one can construct a Hilbert space which admits a strongly continuous irreducible unitary representation of \(\grp{L}_{\mc Q_{i^0}}\). An irreducible representation of the full BMS group is obtained by considering a direct integrable of these Hilbert spaces over the orbit space $\ms{O}_{\mc{Q}_{i^{0}}}$. Since states in these representations have continuous distributions in supermomentum over the orbit space, states with definite supermomentum are improper states in this construction. 

We first recall the case of zero memory where McCarthy's representations correspond to the massless Poincaré representations of Wigner \cite{Wigner}. The vacuum sector of the standard Fock space is the trivial invariant representation of the Poincaré group. The $1$-particle sector is the direct sum of two irreducible massless representations with helicity $\pm 2$ (see also \cite{asymp-quant}) and the $n$-particle sectors are obtained by tensor products of the $1$-particle sectors.  A theorem by Mautner \cite{Mautner} guarantees that any reducible representation of a locally compact group --- such as \(\Fock_0\) --- is also decomposable into direct integrals of the irreducible ones though the explicit decomposition for multi-particle states is rather complicated (see \cite{Lomont,Moussa,Lomont-Moses,schaaf,Barut:1986dd}). The $n$-particle momentum eigenstates are improper states both on $\Fock_{0}$ as well in the group theoretic sense where they are tensor products of a $1$-particle improper basis. Similarly, in the non-zero memory case, the BMS particles are also improper states. Indeed, \(\ket{\ground; \Delta}\) and \(\ket{p_1;\Delta}\) can be considered as improper states in McCarthy's irreducible representations with supermomentum eigenvalues from \cref{thm:eigenstates} and $\ket{p_{1},\dots,p_{n};\Delta}$ can be expressed (non uniquely) as the tensor product of BMS $1$-particle states. 

What does this all mean for scattering theory? While, in the zero memory Fock space, there is a direct compatibility between irreducible representations of quantum fields and of the symmetry group, this compatibility appears to {\em fail} in the case of non-zero memory. Indeed, it follows from \ref{cor:infinite-bms-particles} that any proper state of the field algebra is supported on an infinite number of BMS particles where the different supermomenta are not related by Lorentz transformations. So a proper state in McCarthy's representations is not a proper state on the asymptotic field algebra. Furthermore, the irreducible representations of the field algebra are the memory Fock spaces $\Fock_{\Delta}$ which, as emphasized in \cref{sec:mem-fock}, do {\em not} admit a representation of the full BMS group. While representations of the fields together with the BMS group can be constructed \cite{memory-orbits}, Mautner's theorem does not apply since the BMS group is not locally compact. Therefore, it does not appear to be useful -- or necessary! -- to consider irreducible representations of the BMS group or field algebra separately as is done in the case of zero memory. Instead, the approach taken in this paper is to simply consider states on the full algebra of quantum fields and BMS charges (see \cref{subsec:alg}). 
 \end{remark}

\section{Infrared finite scattering theory: Amplitudes and soft theorems}
\label{sec:amplitudes}
In this section, we define a scattering theory based entirely on the Heisenberg evolution of the fields which thereby allows for transitions between ``in'' and ``out'' states with memory. While, this ``superscattering map'' $\$$ is manifestly well-defined, the fundamental properties and symmetries of $\$$ are relatively obscure. To remedy this, we recast the map $\$$ in terms of ``superscattering amplitudes''. While the superscattering map is sufficient to define a well-defined scattering theory, in order to express this map in terms of amplitudes, one must suitably ``parameterize'' the space of states which arise in scattering theory. To this end, we must generalize the notion of asymptotic completeness used to define traditional $S$-matrix elements to accommodate states with memory.  In \cref{subsec:IRfinitegencomp} we give the general definition of the superscattering map $\$$ and formulate a ``generalized asymptotic completeness'' which, roughly speaking, states that all states that arise in scattering theory can be expressed as superpositions of states in the collection of memory Fock spaces $\Fock_{\Delta}$ defined in \cref{sec:mem-fock}. Using this generalized asymptotic completeness as well as the BMS particle basis constructed in \cref{sec:rigged}, we then formulate $\$$-amplitudes in \cref{subsec:IRamp} and derive an analog of the Weinberg soft graviton theorem for superscattering amplitudes which implies that these amplitudes are well-defined in the infrared. In \cref{subsec:colldiv} we consider the case of theories with collinear divergences. 

\subsection{Scattering theory with massless fields and generalized asymptotic completeness}
\label{subsec:IRfinitegencomp}

In the usual formulation of scattering theory, one assumes that all in/out-states lie in the standard Fock space $\Fock_{0}$ and the $S$-matrix is then defined as the unitary map $\op{S}:\Fock_{0}\to \Fock_{0}$ which evolves an in-state to an out-state in the standard (zero memory) Fock space. As mentioned in the introduction, for scattering with any massless fields the out-state is generically not an element of $\Fock_{0}$ and will be a superposition of states which lie in the collection of memory Fock spaces $\Fock_{\Delta}$. Since the memory is not conserved, any Hilbert space constructed out of a collection of the (uncountably many) $\Fock_{\Delta}$ will generally not scatter into itself under evolution  --- i.e., all elements of the `in'' Hilbert space will generically not scatter into a similarly constructed out-Hilbert space. Using the conserved charges of the theory, one can attempt to significantly restrict the space of ``in'' and ``out'' states by demanding that they are eigenstates of the conserved charges. In QED, this strategy was successfully applied by Faddeev and Kulish \cite{Kulish:1970ut} to obtain a in/out-Hilbert space of states with well-defined energy and angular momentum which scatters into itself. However, it was proven in \cite{PSW-IR} that the analogous construction does not yield a large enough class of physical states in QED with massless charged fields, Yang-Mills theories, and quantum gravity. Indeed, in quantum gravity, the analogous Faddeev-Kulish construction is to construct eigenstates of the supertranslation charges $\op{\mc{Q}}_{i^{0}}(f)$ and it was proven that the only such eigenstate is the vacuum state.\footnote{The proof of this statement is given in theorem 1 of \cite{PSW-IR}. A related but distinct result is that the BMS $0$-particle states $\ket{\ground;\Delta}$ defined in \cref{sec:bms-particles} are non-normalizable states.} It was further argued in \cite{PSW-IR} that no other preferred Hilbert space of states exists in more general quantum field theories and quantum gravity. 

How should scattering theory be defined in the absence of a preferred Hilbert space? Most of the assumptions of the standard formulation of scattering theory can, in fact, be carried over without much change; it is only the assumption of asymptotic completeness that needs to be revisited as we argue next. We first note that the asymptotic algebras $\Alg_{\inn}$ and $\Alg_{\out}$ are well-defined. These algebras include the asymptotic fields as well as the generators of any symmetry group (e.g. large gauge transformations, Poincaré or BMS symmetries). These algebras were defined in \cite{asymp-quant,PSW-IR} for QED with massive or massless charged sources, Yang-Mills fields as well as in quantum gravity (see \cref{sec:asymp-quant}). Furthermore, as reviewed in \cref{subsec:alg}, states are simply defined as positive linear maps on the algebra and are uniquely specified by their $n$-point correlation functions on the algebra. Therefore, there is no problem in defining in-states $\Psi_{\inn}$ on $\Alg_{\inn}$ or out-states $\Psi_{\out}$ on $\Alg_{\out}$ with well-defined (super)momentum and angular momentum with or without memory. An explicit construction of a large supply of such physical states was provided in \cite{memory-orbits}.

Given that the asymptotic in/out algebras are well-defined, the scattering can be described by evolving the field operators in the in-algebra to the future using the Heisenberg equations of motion. The Heisenberg evolution of the interacting field algebra can be formulated rigorously, including the regularization and renormalization of the ultraviolet divergences, in terms of the ``retarded products'' of the free field algebra; see \cite{YangFeld_1950,Kallen:1950uha,Haag:1955ev,Hollands:2002ux,Duetsch:2004dd,Hollands:2014eia}. These Heisenberg evolution equations then allow us to write any out-operator $\op{O}_\out\in \Alg_{\out}$ in terms of in-operators. That is, there exists an \emph{evolution map} \(\mc E: \Alg_\out \to \Alg_\inn\), which preserves the algebra structure, i.e., for all out-operators \(\op O_\out \in \Alg_\out\) we have \(\mc E[\op O_\out] \in \Alg_\inn\) and satisfies
\be\label{eq:evolution}
    \mc{E}[\op O_{1,\out} +\op O_{2,\out}] = \mc{E}[\op O_{1,\out}] +\mc{E}[\op O_{2,\out}] &\eqsp \mc E[\op O_{1,\out} \op O_{2,\out}] = \mc E[\op O_{1,\out}] \mc E[\op O_{2,\out}]  \, \\ 
    \mc{E}[\op{1}] = \op{1} \eqsp &\textrm{and } \quad \mc E[\op O_\out]^* = \mc E[\op O_\out^*]\,.
\ee
When standard asymptotic completeness holds --- e.g. for a theory with only massive fields --- \cref{eq:evolution} implies that the evolution map \(\mc E\) can be represented on the common Hilbert space of in/out states by a unitary operator \(\op S\), the \(S\)-matrix, as
\be\label{eq:E-S-relation}
&\mc E [\op{O}_\out] = \op S^\dagger \op{ O}_\inn \op S \eqsp \quad &&\textrm{(standard asymptotic completeness).}
\ee
In this case, the evolution of the asymptotic operators is equivalent to the existence of an \(S\)-matrix in the sense of Heisenberg \cite{YangFeld_1950,Kallen:1950uha}.

Next, we use the evolution map \(\mc E\) on operators to define the scattering of asymptotic algebraic states as follows. Given any algebraic in-state $\Psi_{\inn}$ the corresponding out-state $\Psi_{\out}$ is defined by its correlation functions for all out-operators \(\op O_\out\) by\footnote{We note that the out-correlation functions can be computed using \cref{eq:evolution2} purely from the knowledge of the in-state \(\Psi_\inn\) and the Heisenberg evolution map \(\mc E\). This is closer to the spirit of the ``in-in formalism'' \cite{Keldysh:1964ud,Kadanoff-Baym,Weinberg:2005vy} used for non-equilibrium systems and cosmological perturbations rather than the ``in-out formalism'' usually used to compute scattering amplitudes.}
\begin{equation}
\label{eq:evolution2}
\Psi_{\out}(\op{O}_{\out}) = \Psi_{\inn}(\mc{E}[\op{O}_{\out}])
\end{equation}
Thus, \cref{eq:evolution2} defines the superscattering map on algebraic states 
\begin{equation}
\label{eq:superscattering}
\Psi_{\out} = \$ \Psi_{\inn}
\end{equation}
Similar to the evolution map of operators, when the in/out states are assumed to live in the same Hilbert space, the superscattering map in \cref{eq:superscattering} reduces to the standard \(S\)-matrix notion of scattering. To see this, let the in/out states be assumed to live in the zero memory Fock space \(\Fock_{0,\out} \cong \Fock_{0,\inn} \cong \Fock_0\), and let \(\ket{\Psi}\) be a vector in this Hilbert space. This vector defines an algebraic in-state by \(\Psi_\inn(\op O_\inn) \defn \braket{\Psi | \op O_\inn | \Psi}\) for all \(\op O_\inn \in \Alg_\inn\). Since in this framework, the operators are evolved using the equations of motion, the state vector \(\ket{\Psi}\) will remain fixed. The assumption of asymptotic completeness and \cref{eq:E-S-relation} imply that the correlation functions of the out-operators define the algebriac out-state \(\Psi_\out\) by
\be
    \Psi_\out(\op O_\out) = \braket{\Psi | \op S^\dagger \op O_\inn \op S | \Psi} = \Psi_\inn(\mc E[\op O_\out])\eqsp &&\textrm{(standard asymptotic completeness).}
\ee
Thus, when asymptotic completeness holds, the evolution given by the \(S\)-matrix is precisely equivalent to the one defined on algebraic states by \cref{eq:evolution2,eq:superscattering}. However when the in/out states live in inequivalent Hilbert spaces, \cref{eq:evolution2,eq:superscattering} define the scattering map directly in terms of algebraic states without recourse to any pre-selected Hilbert space.

In summary, the evolution map \(\mc E\) --- and consequently also $\$$ --- can be computed to any order in perturbation theory using the Heisenberg evolution equations. When, asymptotic completeness holds, the evolution map is described by a unitary operator on a single separable Hilbert space and the above formulation of scattering is equivalent to the usual \(S\)-matrix formulation. However, for massless fields the \(S\)-matrix formulation leads to IR divergences precisely because the assumption of asymptotic completeness is false. Indeed, the in-state $\Psi_{\inn}$ can be chosen to be an element of the standard Fock space $\Fock_{0,\inn}$ --- though one is not required to make this choice --- however the out-state $\$ \Psi_{\inn}$ will generically have memory and will not lie in $\Fock_{0,\out}$. Nevertheless, we emphasize that, in contrast to the \(S\)-matrix, the \(\$\) map is manifestly infrared finite since no choice of Hilbert space for the ``in'' or ``out'' states is made in the construction. Therefore, while the standard $S$-matrix cannot be defined, the superscattering map $\$$ is well-defined for general quantum field theories and quantum gravity.

Next we consider some of the general properties of the superscattering map. The basic properties which follow from the definition are:
\begin{enumerate}[label=(S.{\Roman*})]
\label{$prop12}
\item (Convex linear)~$\$[\lambda\Psi_{1,\inn} + (1-\lambda)\Psi_{2,\inn}] = \lambda \$\Psi_{1,\inn} + (1-\lambda)\$\Psi_{1,\inn}$ for any $\Psi_{1,\inn}$ and $\Psi_{2,\inn}$ and \(0 \leq \lambda \leq 1\).  \label{S1}
\item (Conservation of probability)~ $\$\Psi_{\inn}(\op{1}) = \Psi_{\inn}(\op{1})$. \label{S2}
\end{enumerate}
Since the convex linear sum of two normalized algebraic states yields an algebraic state, property \ref{S1} simply expresses the ``linearity'' of the superscattering map. Property \ref{S2} follows from \cref{eq:evolution} and expresses the fact that the state remains normalized under evolution. Furthermore, if the initial and final asymptotic algebras are both defined on Cauchy surfaces for the evolution of any fields then the map $\$$ will map pure states to pure states.\footnote{An algebraic state is called ``pure'' if it cannot be expressed as the (convex) sum of other algebraic states.} Thus for any well-posed quantum field theory or perturbative quantum gravity on a (globally hyperbolic) asymptotically flat spacetime the map $\$$ will have the additional property that
\begin{enumerate}[label=(S.{\Roman*})]
\setcounter{enumi}{2}
\label{$prop3}
\item (No information loss) If $\Psi_{\inn}$ is a pure state then $\$\Psi_{\inn}$ is also a pure state. \label{S3}
\end{enumerate}
Whether this property holds in full non-perturbative quantum gravity is the subject of the black hole information paradox (see, e.g., \cite{Unruh:2017uaw,Marolf:2017jkr}) and is outside the scope of this paper. 

The last property of this map concerns conservation laws. Quite generally, the algebras $\Alg_{\inn}$ and $\Alg_{\out}$ contain the generators of the asymptotic symmetry group. In QED and Yang-Mills theories the asymptotic symmetry group corresponds to the Poincaré group together with the group of large gauge transformations. The relevant generators correspond to the energy-momentum, angular momentum, and large gauge charges at spatial infinity. In quantum gravity, the relevant symmetry group is the BMS group and their generators correspond to the BMS charges (see \cref{subsec:alg}). For definiteness, we will focus on the formulation of conservation of the BMS charges in this framework but analogous statements hold for more general quantum field theories. In quantum gravity, the in supertranslation charges $\op{\mc{Q}}^{\inn}_{i^{0}}(f)$ of the in-algebra $\Alg_{\inn}$ are determined by the ``in'' radiative data (see \cref{eq:QJ-mem-op}) and, similarly, the out supertranslation charge $\op{\mc{Q}}^{\out}_{i^{0}}(f)$ is expressed in terms of the ``out'' data. Classically, the in/out supertranslation charges can be expressed as a function $\mc{Q}^{\inn}_{i^{0}}(\hat{p})$ and $\mc{Q}^{\out}_{i^{0}}(\hat{p})$ where $\hat{p}$ corresponds to the angular directions of the (future-directed) null generators of $\scri^{-}$ and $\scri^{+}$ respectively. Conservation of these supertranslation charges corresponds to a matching $\mc{Q}^{\inn}_{i^{0}}(f)=\mc{Q}^{\out}_{i^{0}}(f)$ for any $f$.\footnote{If one instead uses globally defined spherical coordinates as opposed to the angular directions of the null momenta then the BMS charges match by an antipodal reflection on $\bb{S}^{2}$.} Similarly, the Lorentz charges also match as $\mc{Q}^{\inn}_{i^{0}}(X) = \mc{Q}^{\out}_{i^{0}}(X)$. In the quantum theory, the corresponding matching of a general out BMS charge is  $\op{\mc{Q}}^{\out}_{i^{0}}(f,X) = \op{\mc{Q}}_{i^{0}}^{\out}(f)+\op{\mc{Q}}_{i^{0}}^{\out}(X)$ to an in BMS charge $\op{\mc{Q}}^{\inn}_{i^{0}}(f,X)$ is simply given by 
\begin{equation}
\label{eq:chargecons}
\mathcal{E}[\op{\mc{Q}}^{\out}_{i^{0}}(f,X)]  = \op{\mc{Q}}^{\inn}_{i^{0}}(f,X).
\end{equation}
By \cref{eq:evolution2,eq:superscattering} this yields the following constraint on the superscattering map 
\begin{enumerate}[label=(S.{\Roman*})]
\setcounter{enumi}{3}
\item (Conservation of BMS charges) The correlation functions of the BMS charge $\Psi_{\inn}$ with an insertion of any arbitrary operator appropriately ``match'' to correlation functions of the BMS charge $\$\Psi_{\inn}$ with the evolved operators subject to \cref{eq:chargecons}. That is, the correlation functions satisfy \label{S4}
\begin{equation}
\Psi_{\inn}(\op{\mc{Q}}^{\inn}_{i^{0}}(f,X)\mathcal{E}[\op{O}_{\out}])= \$\Psi_{\inn}(\op{\mc{Q}}^{\out}_{i^{0}}(f,X)\op{O}_{\out})\quad  \textrm{ for all $\op{O}_{\out}\in \Alg_{\out}$}
\end{equation}
\end{enumerate}
Analogous conservation laws can be similarly formulated for any conserved quantities in scattering for any general quantum field theory. 

Thus, the superscattering map $\$$ defines a well-defined scattering theory and satisfies the general properties \ref{S1}--\ref{S4}. However, more fine-grained details of the scattering theory are relatively opaque in this framework. In standard scattering theory, the full content of the $S$-matrix is encoded in the scattering amplitudes. In addition to defining the scattering theory, these amplitudes allow one to study the full details of the scattering theory such as analyticity, crossing symmetry, Ward identities, and the Weinberg soft theorem which is of particular relevance to this paper. Indeed, it is not a priori clear that any of these notions make any sense without a preferred Hilbert space of states. 

The key ingredient in defining ``superscattering amplitudes'' is to broaden the notion of asymptotic completeness used in defining traditional scattering amplitudes. In standard scattering amplitudes, one assumes that one can decompose any out-state in terms of momentum states of the (zero memory) Fock space $\Fock_{0}$. This assumption is, of course, blatantly false for the scattering with any massless fields. Indeed a generic scattering state will have memory and, in \cref{sec:mem-fock}, we constructed an uncountably infinite number of memory Fock representations $\Fock_{\Delta}$ which each contain states with memory $\Delta_{AB}$. Therefore a straightforward generalization of asymptotic completeness is to allow the in/out states to be a superposition of states in the memory Fock spaces. Thus, we assume that the states in the Fock spaces $\Fock_{\Delta}$ suitably ``span'' the possible states which arise in scattering. More precisely, given any (``in'' or ``out'') state $\Psi$, we assume that there exists a family of vectors $\ket{\psi(\Delta)}\in \Fock_{\Delta}$ and a measure $\mu_{\Psi,\Delta}$ such that 
\begin{equation}
\label{eq:memdecomp}
\Psi(\op{O}) = \int_{\orbit_{\Psi}}d\mu_{\Psi,\Delta}\braket{\psi(\Delta)|\op{O}|\psi(\Delta)} \text{ for all } \op O \in \Alg
\end{equation}
where $\orbit_{\Psi}$ is a subset of the square-integrable memories which may depend on the chosen state \(\Psi\). A large class of states on $\Alg$ satisfying \cref{eq:memdecomp} were explicitly constructed in \cite{memory-orbits}. Indeed, given any state $\Psi$ satisfying \cref{eq:memdecomp}, the Gel'fand-Naimakr-Segal (GNS) construction with respect to $\Alg$ yields a dense set of Hadamard states satisfying \cref{eq:memdecomp}. Therefore, there is an abundant supply of states which satisfy this condition. The generalized asymptotic completeness is then the assumption that all states that arise in scattering theory can be decomposed as in \cref{eq:memdecomp}. We note that this decomposition does not require one to choose a preferred Hilbert space. Indeed, given any two states $\Psi$  and $\Psi^{\prime}$ satisfying \cref{eq:memdecomp}, their respective GNS Hilbert spaces will generically be unitarily inequivalent. We summarize the above described generalization of asymptotic completeness as follows:
\begin{ass}[Generalized asymptotic completeness]
\label{genasympcomp}
Given any state $\Psi$ on $\Alg$, there exists a measure $\mu_{\Psi,\Delta}$ on a subspace of memories $\orbit_\Psi \subseteq L^{2}(\bb{S}^{2})$ and a family of vectors $\ket{\psi(\Delta)}\in \Fock_{\Delta}$ for all $\Delta_{AB} \in \orbit_\Psi$ which satisfy \cref{eq:memdecomp}.
\end{ass}

For perturbative scattering in QED and quantum gravity, the ``non-Fock'' behavior of the state appears to be entirely due to the fact that the radiation field has memory at low-frequencies \cite{Bloch_1937,Weinberg:1965}. Therefore, assumption \ref{genasympcomp} 
seems to be a valid assumption in these cases. However, as we will explain in \cref{subsec:colldiv}, assumption \ref{genasympcomp} will {\em not} hold in theories with collinear divergences such as massless QED and Yang-Mills theories. Thus, while the superscattering map $\$$ is well-defined, it seems one must further generalize the notion of asymptotic completeness to obtain well-defined amplitudes in these theories. 

\subsection{Superscattering amplitudes}
\label{subsec:IRamp}
As described in the introduction, the standard $S$-matrix is assumed to be a unitary operator $\op{S}:\Fock_{0} \to \Fock_{0}$ on the standard Fock space. To analyze the properties of this map it is convenient to repackage the $S$-matrix in terms of amplitudes which are components of $\op{S}$ in a complete basis of $\Fock_{0}$. Any basis of $\Fock_{0}$ can be used to express these amplitudes, but the $n$-particle momentum basis is particularly convenient 
\be
\label{eq:Smatrixamp}
    \braket{p_1^\out,\ldots,p_n^\out | \op S | p_1^\inn,\ldots,p_m^\inn}
\ee
since, in this basis, momentum conservation is straightforward to impose. Using the LSZ reduction formulas \cite{LSZ}, these amplitudes can be expressed in terms of time-ordered products of local field operators and can be computed --- to any finite order in perturbation theory --- using Feynman diagrams. However, for scattering with massless fields, the evolution of quantum fields cannot generically be defined on a single separable Hilbert space, and the amplitudes given by \cref{eq:Smatrixamp} are not normalizable in $\Fock_{0}$. 

In the previous subsection, we defined a superscattering map $\$$ which does not make any a priori choice of in/out Hilbert space and is manifestly infrared finite. How should one define superscattering amplitudes in the framework? By assumption \ref{genasympcomp}, any state $\Psi$ can be decomposed into states in the memory Fock spaces. Furthermore, in \cref{sec:rigged} we defined the improper BMS \(n\)-particle basis on each Fock space $\Fock_{\Delta}$ which generalizes the plane wave basis of $\Fock_{0}$. Therefore, any state $\Psi$ can be decomposed in terms of BMS particles and scattering theory can be fully expressed in terms of $\$$-amplitudes given by 
\be\label{eq:algebraic-amplitude}
    \braket{p_1^\out,\ldots,p_n^\out; \Delta^\out | \,\$\, | p_1^\inn,\ldots,p_m^\inn; \Delta^\inn} \,,
\ee
where \(\ket{p_1^\inn,\ldots,p_m^\inn; \Delta^\inn}\) are viewed as an improper basis for the algebraic in-state \(\Psi_\inn\) and \(\ket{p_1^\out,\ldots,p_n^\out; \Delta^\out}\) are viewed as an improper basis for the algebraic out-state \(\Psi_\out\). Note that because we have suppressed the spherical tensor indices in the notation for the BMS \(n\)-particle basis, this \(\$\)-amplitude has \(n\) pairs of lower indices \(A_1B_1\ldots A_nB_n\) from the ``out'' BMS particles and similarly, \(m\) pairs of upper indices from the ``in'' BMS particles. While familiar methods such as LSZ reduction and Feynman diagrams do not straightforwardly apply to $\$$-amplitudes since these methods are adapted to cases where all states lie in the standard Fock space, recasting the scattering map in terms of amplitudes enables one to investigate other fine-grained details of the scattering theory in a manifestly well-defined manner. In the following, we investigate the implications of invariance of $\$$ under asymptotic symmetries.

The conservation of the BMS charges in quantum scattering theory was originally conjectured by Strominger \cite{Strominger:2013jfa} and is formulated within the context of superscattering theory in property \ref{S4} of the previous subsection. In particular, we restrict consideration to the conservation of supermomentum. The BMS particles have a well-defined action of the supermomentum and property \ref{S4} implies that the supermomentum operator must commute with the superscattering map when inserted into the $\$$-amplitude in \cref{eq:algebraic-amplitude}. Using the fact that the BMS particle basis are eigenstates of the supermomentum (\cref{thm:eigenstates})
\be
    0 &= \braket{p_1^\out,\ldots,p_n^\out; \Delta^\out | \lb[ \op{\mc Q}_{i^0}(f), \,\$\, \rb] | p_1^\inn,\ldots,p_m^\inn; \Delta^\inn} \\
    & = \lb[ \sum_{i=0}^n \omega_i^\out f(\hat p_i^\out) + \Delta^\out(f) - \sum_{i=0}^m \omega_i^\inn f(\hat p_i^\inn) - \Delta^\inn(f) \rb] \\
    &\qquad\qquad \times \braket{p_1^\out,\ldots,p_n^\out; \Delta^\out | ~\$~ | p_1^\inn,\ldots,p_m^\inn; \Delta^\inn}
\ee
for all the BMS particle basis and all supertranslation functions \(f\). Thus, the $\$$-amplitude vanishes unless we have that 
\begin{equation}
\label{eq:supertranscon}
\sum_{i=0}^{n}\omega_{i}f(\hat{p}^{\out}_{i}) + \Delta^{\out}(f) = \sum_{i=0}^{m}\omega_{i}f(\hat{p}^{\inn}) + \Delta^{\inn}(f)
\end{equation}
For $f$ supported on $\ell=0,1$ spherical harmonics, the memory vanishes and \cref{eq:supertranscon} simply implies conservation of $4$-momentum 
\begin{equation}
\sum_{i=0}^{n} p_{i}^{\out,\mu} = \sum_{i=0}^{m}p_{i}^{\inn,\mu}
\end{equation}
where we recall that $p^{\mu}$ has been identified with a null vector in Minkowski spacetime (see \cref{eq:p-defn}). For the higher harmonics we may ``unsmear'' the supertranslation function to obtain 
\be\label{eq:charge-conservation-unsmeared}
    \sum_{i=0}^n \omega_i^\out \delta_{\bb S^2}(\hat p, \hat p_i^\out) \big\vert_{\ell \geq 2} + \ms D^A \ms D^B\Delta^\out_{AB}(\hat p) &= \sum_{i=0}^m \omega^\inn_i \delta_{\bb S^2}(\hat p, \hat p_i^\inn) \big\vert_{\ell \geq 2} + \ms D^A \ms D^B \Delta^\inn_{AB}(\hat p)
\ee
As noted in \cite{Ashtekar:2018lor}, this is an extremely stringent condition on the \(\$\)-amplitudes.\footnote{\Cref{eq:charge-conservation-unsmeared} was obtained in \cite{Ashtekar:2018lor} under the assumption of the existence of an infrared finite scattering theory as well a basis of eigenstate of the supermomentum on $\Fock_{\Delta}$. However, neither the scattering theory nor the basis of eigenstates was defined.} For instance, if we assume a priori that the ``in'' and ``out'' states have the same memory i.e. \(\Delta^\out = \Delta^\inn\), the \(\$\)-amplitude vanishes except when \(m=n\) and \((p_1^\out,\ldots,p_n^\out) = (p_1^\inn,\ldots,p_n^\inn)\) i.e. the scattering of BMS particles is trivial in the sense that the in-particles are simply permuted under scattering. For non-trivial scattering, \cref{eq:charge-conservation-unsmeared} implies that the change in memory between the in/out-state cannot be smooth, however, the memory is still square-integrable on $\bb{S}^{2}$. To see this, we note that, given an in-memory the solutions to \cref{eq:charge-conservation-unsmeared} can be obtained by summing solutions $\Delta^{\out}_{AB}(\hat{p};p_{i})$ which satisfy 
\begin{equation}
\label{eq:DDelta}
    \ms{D}^{A}\ms{D}^{B}\Delta_{AB}^{\out}(\hat{p};p_{i}) = \omega_{i}\delta_{\bb{S}^{2}}(\hat{p},\hat{p}_{i})\big\vert_{\ell\geq 2}
\end{equation}
To solve this equation, it is convenient to introduce stereographic coordinates on $\bb{S}^{2}$. Each point $\hat{p}\in \bb{S}^{2}$ can be parameterized by complex coordinates $\hat{p}=(z,\bar{z})$. In these coordinates the metric and area element on $\bb{S}^{2}$ are given by 
\begin{equation}
q_{AB}\defn 2P^{-2}dzd\bar{z}, \quad d^{2}\hat{p} \defn P^{-2}d^{2}z
\end{equation}
where $d^{2}z$ is the natural area element on the complex plane and $P(z,\bar{z}) \defn \frac{1}{\sqrt{2}}(1+|z|^{2})$. We choose a complex null basis on $\bb{S}^{2}$ by 
\begin{equation}
m_{A}\defn P^{-1}dz,\quad \bar{m}_{A} = P^{-1}d\bar{z}
\end{equation}
such that $m_{A}m^{A}=\bar{m}_{A}\bar{m}^{A} = 0$ and $m^{A}\bar{m}_{A}=1$, and the metric takes the form of $q_{AB}=2m_{(A}\bar{m}_{B)}$ in this basis. We may decompose the memory tensor in this basis into the components 
\begin{equation}
\Delta_{AB} = \Delta(z,\bar{z})m_{A}m_{B}+ \bar{\Delta}(z,\bar{z})\bar{m}_{A}\bar{m}_{B}
\end{equation}
where $\Delta(z,\bar{z})$ is a complex function.\footnote{$\Delta(z,\bar{z})$ has conformal weight $w=-1$ and spin weight $s=-2$ (see, e.g., appendix A of \cite{memory-orbits} for a definition of these weights)} In these coordinates, \cref{eq:DDelta} becomes
\begin{equation}
\label{eq:DDPDelta}
P^{2}\frac{\partial}{\partial \bar{z}}\bigg(P^{2}\frac{\partial}{\partial \bar{z}}(P^{-2}\Delta^{\out})\bigg) = \frac{\omega_{i}}{2}\delta_{\bb{S}^{2}}(z-z_{i},\bar{z} - \bar{z}_{i})\big\vert_{\ell \geq 2}. 
\end{equation}
Thus we seek a distributional solution (or Green's function) to \cref{eq:DDPDelta} such that the solution vanishes when \cref{eq:DDPDelta} is smeared with an $\ell=0,1$ spherical harmonic. Such Green's functions were obtained by \cite{Green-edth} and are given by\footnote{Note that the stereographic coordinate \(\zeta\) used in \cite{Green-edth} corresponds to our \(\bar z\). So our functions of spin weight \(s\) correspond to spin weight \(-s\) in \cite{Green-edth}.\label{fn:coord-convention}}
\be\label{eq:mem-soln}
    \Delta^{\out}(z,\bar z;p_{i}) = \frac{\omega_i}{4\pi}\, \frac{\bar z - \bar z_i}{z - z_i} \frac{(1 + z_i \bar z)^2}{(1 + z \bar z) (1 + z_i \bar z_i)}
\ee
The general solution to \cref{eq:DDelta} is obtained by summing \cref{eq:mem-soln} over all possible ``in'' and ``out'' BMS particle momenta appearing in \cref{eq:algebraic-amplitude} together with the term with the ``in'' memory $\Delta^{\inn}(z,\bar{z})$. It is straightforward to check that, despite the fact that \cref{eq:mem-soln} is not smooth, this solution is square-integrable on the sphere and therefore, as expected, the radiation fields that arise in scattering have finite energy.\footnote{Note that the absolute value of the solution in \cref{eq:mem-soln} is \(O(1)\) as one approaches the singular point, \(z \to z_i\). It was erroneously stated in \cite{PSW-IR} that it diverges logarithmically; but the rest of the conclusions in that paper still hold.}

The above relations control the low-frequency behavior of $\$$-amplitudes. For the standard $S$-matrix amplitudes, the low-frequency behavior is captured by the Weinberg soft graviton theorem which states that, in the limit as one of the frequencies in the amplitude vanishes, the $S$-matrix elements ``factorize'' into a divergent term and a term that depends entirely on other momenta \cite{Weinberg:1965}. Indeed, the divergence that appears in the Weinberg soft theorem is not normalizable in $\Fock_{0}$ and directly implies that (for non-trivial scattering) all $S$-matrix amplitudes are not well-defined in the infrared. We now prove the analog of the Weinberg soft graviton theorem for the infrared-finite $\$$-amplitudes. As we will see this factorization of the $\$$-amplitudes follows directly from the soft factorization of the BMS particles shown in lemma \ref{lem:soft-factor}. Applying this lemma to the $\$$-amplitudes yields 
\be\label{eq:factorization-amplitude}
    \lim_{\omega_n^\out \to 0} &\braket{p_1^\out,\ldots,p_n^\out; \Delta^\out | \,\$\, | p_1^\inn,\ldots,p_m^\inn; \Delta^\inn} \\
    &\qquad\qquad\qquad = \sqrt{\frac{2}{\pi n}}~ \Delta_{A_nB_n}^\out(\hat p_n^\out) \braket{p_1^\out,\ldots,p_{n-1}^\out; \Delta^\out | \,\$\, | p_1^\inn,\ldots,p_m^\inn; \Delta^\inn}
\ee
To express this result in the familiar form of the Weinberg soft graviton theorem we will need to introduce some further notation. Recall that $p$ can be identified with a future-directed null vector $p^{\mu}$ on the positive light cone of Minkowski spacetime (\cref{eq:p-defn} with \(\omega \geq 0\)). Furthermore the angular directions of the null momenta \(\hat p\) are identified with the coordinates \((z,\bar z)\) on \(\bb S^2\) by stereographic projection onto \(\bb C\). Explicitly we have that the Cartesian components of \(p^\mu\) are
\be
    p^\mu = \frac{\omega}{\sqrt{2}} P^{-1} \begin{pmatrix} 1+\abs{z}^2,& z+\bar z,& -i (z- \bar z),& 1-\abs{z}^2 \end{pmatrix}
\ee
We also define the \emph{polarization vectors} relative to \(p\) by
\be\label{eq:polarization-defn}
    \epsilon^\mu(\hat p) &\defn \omega^{-1} P\pd{\bar z} p^\mu = \frac{1}{2}P^{-1} \begin{pmatrix} 0, & 1 - z^2, & i (1+ z^2), & -2 z \end{pmatrix} \\
    \bar\epsilon^\mu(\hat p) &\defn \omega^{-1}P\pd{z} p^\mu = \frac{1}{2}P^{-1} \begin{pmatrix} 0, & 1 -\bar z^2, & -i (1+\bar z^2), & -2\bar z \end{pmatrix} 
\ee
which satisfy
\be
\label{eq:epsilon}
    p^\mu \epsilon_\mu = p^\mu \bar\epsilon_\mu = \epsilon^\mu \epsilon_\mu = \bar\epsilon^\mu \bar\epsilon_\mu = 0 \eqsp \epsilon^\mu \bar\epsilon_\mu = 1.
\ee
With this notation, the solution in \cref{eq:mem-soln} can be rewritten as
\be
\label{eq:outmemWeinberg}
    \Delta^{\out}(z,\bar z;p_{i}) = - \frac{1}{4\pi} \frac{p_i^\mu p_i^\nu \bar\epsilon_{\mu\nu}}{p_i^\mu \hat{p}^\nu \eta_{\mu\nu}} \,,
\ee
where the polarization tensor \(\bar\epsilon_{\mu\nu} = \bar\epsilon_\mu \bar\epsilon_\nu\) is defined relative to \(p\). Using \cref{eq:charge-conservation-unsmeared} and \cref{eq:outmemWeinberg} to express the memory $\Delta_{A_nB_n}^\out(\hat p_n^\out)$ in \cref{eq:factorization-amplitude} in terms of the momenta of the other in/out-particles appearing in the $\$$-amplitude we obtain the infrared-finite soft graviton theorem, summarized below.

\begin{thm}[Infrared finite soft graviton theorem]\label{thm:soft}
Given the conservation of BMS supermomenta (\cref{eq:charge-conservation-unsmeared}) and the soft factorization of the BMS particles (lemma \ref{lem:soft-factor}), the infrared finite \(\$\)-amplitudes satisfy
\be
\label{eq:softamp}
    &\lim_{\omega_n^\out \to 0} \braket{p_1^\out,\ldots,p_n^\out; \Delta^\out | \,\$\, | p_1^\inn,\ldots,p_m^\inn; \Delta^\inn} = S^{(0)}_{A_{n}B_{n}} \braket{p_1^\out,\ldots,p_{n-1}^\out; \Delta^\out | \,\$\, | p_1^\inn,\ldots,p_m^\inn; \Delta^\inn}
\ee
where 
\begin{equation}
\label{eq:S0}
S^{(0)}_{A_{n}B_{n}} = \sqrt{\frac{2}{\pi n}}~ \lb[ \Delta^\inn_{A_nB_n}(\hat p_n^\out) - \frac{1}{4\pi} \lb( \sum_{i=0}^m \frac{p_i^{\inn, \mu} p_i^{\inn, \nu} \bar\epsilon_{\mu\nu}}{p_i^{\inn, \mu} \hat{p}_n^{\out, \nu} \eta_{\mu\nu}} - \sum_{i=0}^{n-1} \frac{p_i^{\out, \mu} p_i^{\out, \nu} \bar\epsilon_{\mu\nu}}{p_i^{\out, \mu} \hat{p}_n^{\out, \nu} \eta_{\mu\nu}} \rb)m_{A_n}m_{B_n} + \text{c.c.}\rb],
\end{equation}
which, in addition to the other $n+m-1$ momenta, depends only on the direction $\hat{p}_{n}$. We recall that the spherical tensor indices in \cref{eq:softamp} are suppressed, $\epsilon_{\mu\nu} = \epsilon_\mu \epsilon_\nu$ is defined relative to $p_{n}^{\mu}$ as in \cref{eq:polarization-defn} and the ``c.c.'' indicates the complex conjugate of the preceding term. A similar result also holds if one of the in-momenta is taken to limit to zero frequency.
\end{thm}
If the in-memory \(\Delta^\inn_{AB} = 0\), \cref{eq:softamp,eq:S0} is the analog\footnote{In this paper we have chosen to quantize the News $N_{AB}$ as opposed to the more conventional quantization of the shear $\sigma_{AB}$. Since, by eq.~\ref{eq:News-defn}, they are related by a time derivative the
coefficient in the soft theorem in eq.~\ref{eq:softamp} differs from the one in \cite{Weinberg:1965} by a factor of $1/\omega_{n}^{\out}$.} of the Weinberg soft graviton theorem for superscattering amplitudes. As expected, we do not get any infrared divergence since we are using the infrared finite \(\$\)-amplitudes which take into account all the memory states. 

In what sense is $\$$-amplitude infrared finite? To clarify this point we will, for simplicity, restrict to incoming states $\Psi_{\inn}$ with $\Delta^{\inn}=0$. Theorem~\ref{thm:soft} illustrates that the infrared behavior of $\$$-amplitudes is essentially equivalent to the infrared behavior of the standard S-matrix element which, in our conventions, is given by $\braket{p_1^\out,\ldots,p_n^\out; 0 | \op{S} | p_1^\inn,\ldots,p_m^\inn; 0 }$. The standard S-matrix element attempts to compute the amplitude for the scattering map $\op{S}:\Fock_{0}\to \Fock_{0}$ between zero memory Fock spaces. However, for non-trivial scattering, memory is generically produced and this is reflected in the fact that the low-frequency behavior of $\braket{p_1^\out,\ldots,p_n^\out; 0 | \op{S} | p_1^\inn,\ldots,p_m^\inn; 0 }$ is not normalizable in $\Fock_{0}$. This is the sense in which conventional S-matrix amplitudes are ``infrared divergent''. In contrast, by eq.~\ref{eq:charge-conservation-unsmeared}  the $\$$-amplitude $\braket{p_1^\out,\ldots,p_n^\out; 0 | \$ | p_1^\inn,\ldots,p_m^\inn; 0 }$ between zero memory plane waves {\em vanishes} for non-trivial scattering which is the physically correct answer. Furthermore, by theorem~\ref{thm:soft}, the infrared behavior of the amplitude $\braket{p_1^\out,\ldots,p_n^\out; \Delta^{\out} | \$ | p_1^\inn,\ldots,p_m^\inn; 0 }$ where $\Delta^{\out}$ satisfies eq.~\ref{eq:charge-conservation-unsmeared} is normalizable in the Fock space $\Fock_{\Delta^{\out}}$. Therefore, the probability of scattering to any out-state (with memory) is well-defined and, in this sense, the $\$$-amplitudes are infrared finite.

The conservation of the Lorentz charges have also been related to a type of soft theorem for the \(S\)-matrix amplitudes \cite{Cachazo:2014fwa}. In the infrared finite setting, as explained in \cref{sec:mem-fock} above, the action of the Lorentz charges on a non-zero memory Fock space is not defined since the memory is not Lorentz-invariant. As emphasized in \cite{memory-orbits} conservation of Lorentz charge is only well-defined if one considers direct integrals over the memory Fock spaces. In particular, the formal frequency expansion of the standard S-matrix amplitudes results in a tower of ``soft theorems'' \cite{Lysov:2014csa,Low:1958sn,Low:1958sn,Hamada:2018vrw,Guevara:2021abz,Freidel:2021dfs,Strominger:2021mtt}. We leave the analysis of these soft theorems --- as well as other properties of interest such as analyticity and crossing symmetry --- in the context of $\$$-amplitudes to future investigation. 

\begin{remark}[Infrared finite soft photon theorem]
\label{rem:IRfinitesoftphoton}
    In QED with massive charged fields, the analog of BMS particle states can be constructed in an identical manner as in \cref{sec:rigged} where these states are now eigenstates of the stress-energy flux of the electromagnetic field. The BMS particle states $\ket{p_{1}\dots p_{n},k_{1}\dots k_{m};\Delta}$ are now labeled the null momenta $p_{i}$ of the electromagnetic field and the timelike momenta $k_{i}$ of the charged field. The BMS particles are also eigenstates of the conserved ``large gauge charges'' $\op{\mc{Q}}_{i^{0}}(\lambda)$ where $\lambda(\hat{p})$ is a function on $\bb{S}^{2}$. Conservation of the large gauge charges in $\$$-amplitudes implies that the out-memory can be expressed in terms of the incoming and outgoing momenta $k_{i}^{\inn/\out}$ of the charged field. The $\$$-amplitude in the limit as the frequency $\omega_{n}$ of the $n$th photon implies an analog of the Weinberg soft photon \cite{Weinberg:1965} for superscattering amplitudes in QED\footnote{In the special case where the in/out memories are chosen such that $\Psi_{\inn/\out}$ both are eigenstates of the electromagnetic charge operator $\op{\mc Q}_{i^{0}}(\lambda)$ (see, e.g., \S~5 of \cite{PSW-IR} for details), the $\$$-amplitudes are equivalent to ``Faddeev-Kulish amplitudes'' which have been explicitly been shown to be well-defined in the infrared \cite{Kulish:1970ut}.}.
\end{remark}

\subsection{Collinear divergences}
\label{subsec:colldiv}
We now consider cases where the above formalism of $\$$-amplitudes does not apply. This occurs for theories where the standard $S$-matrix suffers from collinear divergences such as massless QED or unconfined Yang-Mills theories. In these theories, the ``non-Fock'' behavior of the asymptotic state is not merely due to the emission of soft radiative quanta. Thus, while the superscattering map $\$$ is well-defined in such theories, one must further generalize the asymptotic completeness given by assumption \ref{genasympcomp} to obtain well-defined $\$$-amplitudes. 

It is instructive to see precisely how the assumption of generalized asymptotic completeness fails for theories with collinear divergences. In the following, we consider the case of massless QED. As in the case of QED with massive charges (see remark \ref{rem:IRfinitesoftphoton}), one can straightforwardly define the analog of BMS particles $\ket{p_{1}\dots p_{n},k_{1}\dots k_{m};\Delta}$ with memory $\Delta_{A}$ and the $p_{i}$ and $k_{i}$ are positive frequency null momenta of the electromagnetic field and the massless charged field, respectively. The BMS particles are also eigenstates of the conserved large gauge charge $\op{\mc{Q}}_{i^{0}}(\lambda)$ with eigenvalue given by a sum of terms of the form $q_{i}\lambda(\hat{k}_{i})$ where $\lambda(\hat{k}_{i})$ are functions on $\bb{S}^{2}$ and $q_{i}=\pm q$ is the charge of particle or antiparticle. Conservation of the large gauge charges in $\$$-amplitudes implies that 
\be\label{eq:charge-conservation-massless-QED}
    \sum_{i=0}^n q_i^\out \lambda(\hat k_i^\out) + \Delta^{\EM, \out}(\lambda) &= \sum_{i=0}^m q_i^\inn \lambda(\hat k_i^\inn) + \Delta^{\EM, \inn}(\lambda)
\ee
where $\Delta^{\EM}(\lambda) = \int d^{2}\hat{k}~ \Delta_{A}^{\EM}(\hat{p})\ms{D}^{A}\lambda(\hat{p})$ is the electromagnetic memory. If $\lambda$ is purely an $\ell=0$ spherical harmonic, \cref{eq:charge-conservation-massless-QED} implies conservation of the total Coulomb charge $\sum_{i=0}^{n}q_{i}^{\out} = \sum_{i=0}^{m}q_{i}^{\inn}$. For the higher harmonics, \cref{eq:charge-conservation-massless-QED} yields 
\be
\label{eq:memmassQED}
    \sum_{i=0}^n q_i^\out \delta_{\bb S^2}(\hat p, \hat k_i^\out)\big\vert_{\ell \geq 1} + \ms D^A \Delta^{\EM, \out}_{A}(\hat p) &= \sum_{i=0}^m q_i^\inn \delta_{\bb S^2}(\hat p, \hat k_i^\inn)\big\vert_{\ell \geq 1} + \ms D^A \Delta^{\EM, \inn}_{A}(\hat p)
\ee
As in the previous subsection, one can solve \cref{eq:memmassQED} using stereographic coordinates and decomposing the memory in a basis 
\(\Delta_A^\EM(z,\bar z) = \Delta^\EM(z,\bar z) m_A + \bar \Delta^\EM(z,\bar z) \bar m_A\), where \(\Delta^\EM(z,\bar z)\) is a complex function.\footnote{The function $\Delta^{\EM}(z,\bar{z})$ has conformal weight $w=-1$ and spin weight $s=-1$} We will also denote the $\hat{k}_{i}=(z_{i},\bar{z}_{i})$ as the the angular direction of the $i$-th momentum of the charged field. In these coordinates, solutions to \cref{eq:memmassQED} are linear combinations of solutions of
\be
    P^2 \pd{\bar z} (P^{-1} \Delta^{\EM,\out}) = \frac{q_i}{2}\, \delta_{\bb S^2}(z-z_i,\bar z - \bar z_i)\big\vert_{\ell \geq 1}.
\ee 
The solutions are given by (see \cite{Green-edth})
\be
\label{eq:mlessQEDmem}
    \Delta^{\EM,\out}(z,\bar z;\hat{k}_{i})  = \frac{q_i}{4\pi \sqrt{2}}\frac{1 + z_i \bar z }{z - z_i} = \frac{q_i\, \omega}{4\pi}\, \frac{k^\mu_i \bar\epsilon_\mu}{k^\mu_i p^\nu \eta_{\mu\nu}}.
\ee
\Cref{eq:mlessQEDmem} is singular when the momentum $k^{\mu}$ of the massless charge and the momentum of the soft photon $p^{\mu}$ align and, as opposed to \cref{eq:mem-soln}, this expression is {\em not} square-integrable on $\bb{S}^{2}$. These divergences are known as {\em collinear divergences}. Thus, even if the in-state has finite energy,  --- e.g., the in-state could simply be incoming electrons and positrons with no photons --- the assumption of generalized asymptotic completeness implies that the out-state has memory of the form of \cref{eq:mlessQEDmem} and thus must have {\em infinite} energy.\footnote{These collinear divergences also arise when one attempts to construct a ``preferred Hilbert space'' of definite large gauge charge where the radiation field of these states have infinite energy \cite{PSW-IR,Kapec:2017tkm}. This is a special case of the more general result given by \cref{eq:mlessQEDmem}.} These collinear divergences also arise in standard $S$-matrix amplitudes where analogous divergences arise in (unconfined) Yang-Mills theories in the ``out'' gluon field. The standard way to ``deal'' with these collinear divergences, is to --- in addition to an infrared cutoff --- impose an angular cutoff and compute sufficiently ``inclusive cross-sections'' \cite{K_1962,LN_1964}.  However, just as in the case of this approach in the context of infrared divergences, the actual out-state which arises in scattering theory is not constructed in this approach. 

What is the out-state which arises in such theories? Just as the infrared divergences in the usual \(S\)-matrix amplitudes signaled that the standard Fock space is not sufficient to describe all scattering states, the collinear divergences in massless QED and unconfined Yang-Mills theories indicate that one must further generalize the space of asymptotic states in order to define amplitudes. Indeed, just as in QED with massive charges, in the ``bulk'' the scattering of any massless charged particles will result in the emission of large number soft photons. However, in contrast to standard QED, in massless QED there is no threshold to pair produce into electron-positron pairs. Such pairs will also radiate soft photons and this process presumably occurs indefinitely. An analogous process occurs in Yang-Mills theories --- which is directly related to the formation of ``jets''. While perturbative \cite{Contopanagos:1991yb,Curci:1978kj,Havemann:1985ra,DelDuca:1989jt,Forde:2003jt,Kosower:1999xi,Feige:2014wja,Hannesdottir:2019umk} and heuristic \cite{Gribov:1981jw} analyses of such out-states have been considered, to our knowledge a complete characterization of the ``out'' states --- at a similar level as provided in \cref{sec:mem-fock} in the case of infrared divergences due to soft radiation ---  that arise in such theories has not been completed. We leave this construction as well as the formulation of infrared-finite superscattering amplitudes in massless QED and Yang-Mills theories for future work. 

\acknowledgments
We would like to thank Holmfridur Hannesdottir for helpful discussions. G.S. is supported by the Princeton Gravity Initiative at Princeton University. This work was supported in part by the NSF grant PHY-2107939 to the University of California, Santa Barbara.

\appendix

\section{Approximating BMS particle states by proper states}
\label{sec:approx-eigenstates}

In \cref{sec:bms-particles}, we constructed BMS particle eigenstates $\ket{p_{1}\dots p_{n};\Delta}$ of the ``null memory'' operator $\op{\mc{J}}(f)$ --- and consequently eigenstates of $\op{\mc{Q}}_{i^{0}}(f)$ --- on $\Fock_{\Delta}$. These states are generalizations of the improper plane wave states $\ket{p_{1}\dots p_{n}}$ on $\Fock_{0}$ which formally are the quantum states corresponding to plane wave News tensors at null infinity. In this appendix we prove in lemma \ref{lem:approx} that the BMS particle states can be approximated to arbitrary precision by proper states in $\Fock_{\Delta}$. A by-product of this analysis is that, in a precise sense, the 
BMS $0$-particle state $\ket{\ground;\Delta}$ can be viewed as the limit of a sequence of coherent states where the limiting News tensor of the state is supported only at zero frequency --- the Fourier transform $\nu_{AB}(p)$ is non-vanishing at only $\nu_{AB}(0,\hat{p}) = \sqrt{2/\pi}\Delta_{AB}(\hat{p}$).  The other BMS particles can be viewed as plane wave states above this improper coherent background. 

\begin{lemma}
\label{lem:approx}
There exists a sequence of states \(\ket{C_\nu(\epsilon)} \in \Fock_\Delta\) such that
\begin{enumerate}
    \item \(\ket{C_\nu(\epsilon)}\) is normalizable, \(\norm{ C_\nu(\epsilon) }^2_\Delta < \infty\), for every \(\epsilon > 0\), but \(\lim_{\epsilon \to 0} \norm{ C_\nu(\epsilon)}^2_\Delta = \infty\), unless the memory vanishes, \(\Delta = 0\). That is, the sequence \(\ket{C_\nu(\epsilon)}\) does not converge to a proper state in the Hilbert topology on \(\Fock_\Delta\), except when the memory is zero.
    \item The sequence of states \(\ket{C_\nu(\epsilon)}\) are approximate eigenstates of \(\op{\mc J}(f)\) for the eigenvalue \(\lambda = 0\), i.e., for all supertranslations \(f\),
    \be
    \frac{\norm{ \lb(\op{\mc J}(f) - \lambda \1 \rb) \ket{C_\nu(\epsilon)} }^2_\Delta}{\norm{C_\nu(\epsilon)}^2_\Delta} = O(\epsilon^2) \eqsp \lambda = 0
    \ee
    \item The sequence \(\ket{C_\nu(\epsilon)}\) converges to \(\ket{\ground;\Delta}\) in the dual topology on \(\ms D'_\Delta\), i.e., for all \(\ket{\psi} \in \ms D_\Delta\) we have
    \be
    \lim_{\epsilon \to 0} \braket{C_\nu(\epsilon) | \psi}_\Delta = \braket{\ground;\Delta | \psi}
    \ee
\end{enumerate}
\begin{proof}
For \(\epsilon > 0\) we pick a sequence of News tensors\footnote{In position space this is the sequence of News tensors \(\nu_{AB}(x; \epsilon) = \frac{1}{\pi} \frac{\sin(\epsilon u)}{u} \Delta_{AB}(\hat x)\).}
\be\label{eq:memory-family}
    \nu_{AB}(p;\epsilon) =
    \begin{cases}
    \sqrt{\frac{2}{\pi}}~ \Delta_{AB}(\hat p), & \abs{\omega} \leq  \epsilon \\
    0, &  \abs{\omega}> \epsilon 
    \end{cases}
\ee
Let \(\sigma_{AB}(p;\epsilon)\) be the corresponding shear tensor defined via \(\nu_{AB}(p;\epsilon) = -2i\omega \sigma_{AB}(p;\epsilon)\). Pick any regular vacuum state \(\ket{\s_\nu} \in \Fock_\Delta\) with the corresponding shear \(\sigma_{AB}\).

We construct a sequence of \emph{unnormalized} coherent states \(\ket{C_\nu(\epsilon)} \in \Fock_\Delta\)
\be\label{eq:approx-eigenstate}
    \ket{C_\nu(\epsilon)} &\defn \lb[ \bar A_\nu \bar B(\sigma,\sigma(\epsilon)) \rb] e^{-\half\, i\Omega(\sigma(\epsilon),\sigma)} e^{-\tfrac{i}{8\pi} \op N(\sigma(\epsilon) - \sigma)}  \ket{\s_\nu} \\
    \text{where } B(\sigma,\sigma(\epsilon)) &\defn \exp \lb[ \half\, i \Omega(\sigma,\sigma(\epsilon)) + \tfrac{i}{4\pi} \Re\lb[\nu(\epsilon)(\bar \sigma - \bar \sigma(\epsilon))\rb] + \tfrac{1}{128\pi^2} \norm{\sigma - \sigma(\epsilon)}_0^2 \rb]
\ee
Note that without the factors of \(A_\nu\) and \(B\) this would define a sequence of unit-normalized coherent state in \(\Fock_\Delta\) (see \cref{sec:coherent-states}).

We can compute the norm of \(\ket{C_\nu(\epsilon)}\) directly from \cref{eq:approx-eigenstate} to be
\be
    \norm{C_\nu(\epsilon)}_\Delta^2 = \abs{A_\nu B(\sigma,\sigma(\epsilon))}^2 = \abs{A_\nu}^2 e^{\tfrac{1}{64\pi^2} \norm{\sigma - \sigma(\epsilon)}_0^2}
\ee
Note that for \(\epsilon > 0\), \(\norm{\sigma - \sigma(\epsilon)}_0^2\) is finite. We can estimate its limit as \(\epsilon \to 0\) as follows:
\be
    \tfrac{1}{64\pi^2} \norm{\sigma - \sigma(\epsilon)}_0^2 &= \tfrac{1}{16\pi} \int_{C_+} d^3p~ \omega^{-1} \abs{\nu_{AB} - \nu_{AB}(\epsilon)}^2 \\
    &=\tfrac{1}{16\pi} \int_0^\epsilon d\omega \int_{\bb S^2} d^2\hat p~ \omega^{-1} \abs{\nu_{AB} - \sqrt{\tfrac{2}{\pi}} \Delta_{AB}}^2 + \tfrac{1}{16\pi} \int_\epsilon^\infty d\omega \int_{\bb S^2} d^2\hat p~ \omega^{-1} \abs{\nu_{AB}}^2
\ee
Since \(\nu_{AB}(\omega=0, \hat p) = \sqrt{2/\pi}\Delta_{AB}\) and the memory is square-integrable, the first integral vanishes in the limit as \(\epsilon \to 0\), while the second one diverges as \(\lb(- (1/8\pi^2) \norm{\Delta}^2_{L^2(\bb S^2)} \rb) \ln\epsilon\). So, as \(\epsilon \to 0\) we have
\be
    \norm{C_\nu(\epsilon)}_\Delta^2 \sim \epsilon^{- \tfrac{1}{8\pi^2} \norm{\Delta}^2_{L^2(\bb S^2)}}
\ee
Thus while \(\ket{C_\nu(\epsilon)}\) is a proper state in \(\Fock_\Delta\) for all \(\epsilon > 0\), the limit as \(\epsilon \to 0\) does not converge to a proper state in the Hilbert topology on \(\Fock_\Delta\), except when \(\Delta = 0\).

Next we show that the sequence \(\ket{C_{\nu}(\epsilon)}\) approximates an eigenstate of \(\op{\mc J}(f)\) with zero eigenvalue. From \cref{eq:J-coherent}
\be
    \op{\mc J}(f) \ket{C_\nu(\epsilon)} &= \frac{1}{16\pi} \op a(f\nu(\epsilon))^* \ket{C_\nu(\epsilon)} \\
    \norm{ \op{\mc J}(f) \ket{C_\nu(\epsilon)} }^2_\Delta &= \frac{1}{(16\pi)^2} \braket{ C_\nu(\epsilon) | \op a(f\nu(\epsilon)) \op a(f\nu(\epsilon))^* | C_\nu(\epsilon)}
\ee
Using the commutation relations \cref{eq:ccr} we have
\be
    \frac{\norm{ \op{\mc J}(f) \ket{C_\nu(\epsilon)} }^2_\Delta}{\norm{C_\nu(\epsilon)}^2_\Delta}
    &= \frac{1}{(16\pi)^2} \lb[ 16\pi \int_{C_+}d^3p~\omega f^2(\hat p) \abs{\nu_{AB}(\epsilon)}^2  + \abs{ \int_{C_+} d^3p~ f \abs{\nu_{AB}(\epsilon)}^2  }^2 \rb] = O(\epsilon^2)
\ee
Thus, \(\ket{C_\nu(\epsilon)}\) approximates an eigenstate of \(\op{\mc J}(f)\) with zero eigenvalue.

Thus, far we have seen that \(\ket{C_\nu(\epsilon)}\) does not converge to a proper state in \(\Fock_\Delta\) but approximates an eigenstate of \(\op{\mc J}(f)\) of eigenvalue zero. Now we show that it does indeed converge to \(\ket{\ground;\Delta}\) as an improper state, in the sense of the dual topology of \(\ms D'_\Delta\). From \cref{eq:approx-eigenstate,eq:coherent-state-inp}, we have
\be
    \braket{C_\nu(\epsilon) | \s_\nu} = A_\nu
\ee
which also holds in the limit. Next, for any \(\ket{\psi} \in \ms D_\Delta\) take
\be
    \lim_{\epsilon \to 0} \braket{C_\nu(\epsilon) | \psi} &= \lim_{\epsilon \to 0} \sum_{m=0}^\infty~ \int_{C_+^m}d^3p_1\ldots d^3p_m~ \psi^{A_1B_1\ldots A_mB_m}(p_1,\ldots,p_m) \\
    &\qquad\qquad\qquad \times \braket{C_\nu(\epsilon) | \op a_{A_1B_1}(p_1)^* \cdots \op a_{A_mB_m}(p_m)^* | \s_\nu} \\
    &= \lim_{\epsilon \to 0} A_\nu \sum_{m=0}^\infty~ \int_{C_+^m}d^3p_1\ldots d^3p_m~ \psi^{A_1B_1\ldots A_mB_m}(p_1,\ldots,p_m) \\
    &\qquad\qquad\qquad \times \bar \nu_{A_1B_1}(p_1;\epsilon) \cdots \bar \nu_{A_mB_m}(p_m;\epsilon) \\
    & = 0
\ee
where in the last line we have substituted \(\nu_{AB}(\epsilon)\) with the memory following \cref{eq:memory-family}, and since \(\psi^{A_1B_1\ldots A_mB_m}\) is a test tensor the limit of that integral vanishes. Thus, the limit of \(\ket{C_\nu(\epsilon)}\) reproduces the definition of \(\ket{\ground;\Delta}\) when acting on all states \(\ket{\psi} \in \ms D_\Delta\) and so the limit, in the dual topology on \(\ms D'_\Delta\), is the improper state \(\ket{\ground;\Delta}\).
\end{proof}
\end{lemma}
 
Similarly starting with the proper state
\be
    \tfrac{1}{\sqrt{n!}}\int_{C_+^n} d^3p'_1\cdots d^3p'_n~ F_\epsilon(p'_1,\ldots,p'_n) \op a(p'_1)^* \cdots \op a(p'_n)^*\ket{\nu_\epsilon}
\ee
where \(F_\epsilon(p'_1,\ldots,p'_n)\) is a sequence of smooth functions of compact support which limits to a product of delta functions concentrated at \(p'_i = p_i\) as \(\epsilon \to 0\), gives a limiting sequence which approximates \(\ket{p_1,\ldots,p_n;\Delta}\).

Finally, we note that these approximate sequences of eigenstates play a key role in rigorous analyses of the spectral theorem in terms of projection-valued measures as described in \cite{Reed-Simon}. In  \cref{sec:rigged}, we did not employ such techniques and instead used the rigged Hilbert space formalism to explicitly construct the relevant eigenstates. Nevertheless, the results of this appendix indicate that the BMS particles could be given an interpretation in terms of projection-valued measures as well. 

\section{Inner product on memory Fock spaces in terms of BMS particles}
\label{sec:inp-mem}
As shown in theorem \ref{thm:completeness}, the BMS \(n\)-particle states form a complete set of improper basis for any memory Fock space \(\Fock_\Delta\). Therefore one expects that the inner product on \(\Fock_\Delta\) can be expressed in terms of the ``overlap'' of proper states with the BMS particle states. In this appendix we obtain such a formula for the inner product on any memory Fock space.

Let us first recall, the corresponding formula for zero memory. As is well-known the Fock space \(\Fock_0\) can be written as a direct sum over the \(n\)-particle Hilbert spaces as in \cref{eq:fock-0}. Given \(\ket{\psi}, \ket{\psi'} \in \ms D_0\) their inner product can be expressed as
\be\label{eq:inp-0}
    \braket{\psi' | \psi}_0 = \sum_{n=0}^\infty (16\pi)^{-n} \int_{C_+^n} \lb( \prod_{i=1}^n  d^3p_i \omega_i^{-1} \rb) \braket{\psi' | p_1,\ldots,p_n; 0} \braket{p_1,\ldots,p_n; 0 | \psi}
\ee
where
\be
    \braket{\psi | p_1,\ldots,p_n; 0} \defn \bar{ \braket{ p_1,\ldots,p_n; 0 | \psi }}
\ee
and in the product expression \(\braket{\psi' | p_1,\ldots,p_n; 0} \braket{p_1,\ldots,p_n; 0 | \psi}\) above, all the tensor indices on \(\bb S^2\) are contracted . For simplicity, we have also taken  \(\ket{\ground; 0} = \ket{\s_0} \in \Fock_0\) with \(A_0 = 1\) as explained in \cref{rem:zero-memory-simplify} even though the choice of this constant is not important.

We would like to obtain a similar formula for the inner product on \(\Fock_\Delta\) in terms of the BMS particle states with memory. Note that a simple generalization of \cref{eq:inp-0} will not work since, as shown in corollary \ref{cor:infinite-bms-particles}, the only state in \(\ms D_\Delta\) which is supported on finitely many BMS particles is the zero state. Instead, our strategy to obtain such a formula is to first  ``truncate'' states in \(\ms D_\Delta\) with a frequency cut-off and map them to states in \(\Fock_0\). Then, we use \cref{eq:inp-0} to obtain the inner product on \(\Fock_\Delta\) in the limit as the truncation is removed. We explain this truncation procedure next, the final result and proof are collected in theorem \ref{thm:inp-memory} below.

Let \(\theta_\epsilon(\omega) \defn \theta(\omega-\epsilon)\) be the piecewise-defined Heaviside step function which is \(1\) for \(\omega \geq \epsilon\) and vanishes otherwise. For any regular News tensor \(\nu_{AB}\) with memory \(\Delta_{AB}\) consider the \emph{truncated} News tensor \(\nu_{AB}(p;\epsilon) \defn \nu_{AB}(p)\theta_\epsilon(\omega) \) so that for any \(\epsilon > 0\) the memory of \(\nu_{AB}(\epsilon)\) vanishes. Then, the normalized coherent state \(\ket{\omega_{\nu(\epsilon)}}\) constructed out of \(\nu_{AB}(\epsilon)\) lies in \(\Fock_0\) for any \(\epsilon >0\). For any operator \(\op O\) spanned by products of the creation and annihilation operators smeared with test tensors we have from \cref{eq:aut-defn}
\be\label{eq:correlation-lim}
    \s_\nu(\op O) = \s_0(\mf a_\nu\lb[\op O\rb]) = \lim_{\epsilon \to 0} \s_0(\mf a_{\nu(\epsilon)}\lb[\op O\rb]) = \lim_{\epsilon \to 0} \omega_{\nu(\epsilon)}(\op O)
\ee
We emphasize here that the state \(\ket{\omega_{\nu(\epsilon)}}\) \emph{does not} limit to the state \(\ket{\s_\nu}\) in the Hilbert space topology. It is only the correlation functions of the operator \(\op O\) of the type specified above that have limits. For example, if \(\op O\) were replaced by the memory operator, \cref{eq:correlation-lim} would not hold.

Similarly, for a state \(\ket{\psi} \in \ms D_\Delta\) of the form \cref{eq:nice-state} we can associate with it a truncated state \(\ket{\psi(\epsilon)} \in \Fock_0\) given by the same formula as \cref{eq:nice-state} --- with the same test tensors --- but with \(\ket{\s_\nu}\) replaced by the truncated vacuum \(\ket{\omega_{\nu(\epsilon)}}\). From the explicit expressions \cref{eq:0-plane-wave-defn,eq:pn-nice-state} it is easy to see that
\be\label{eq:pn-trunc}
    \lb( \prod_{i=1}^n \theta_\epsilon(\omega_i) \rb) \braket{p_1,\ldots,p_n; \Delta | \psi } &= \frac{A_\nu}{A_{\nu(\epsilon)}} \braket{p_1,\ldots,p_n; 0 | \psi(\epsilon)} \\
    &= \frac{A_\nu}{e^{-\tfrac{1}{128\pi^2} \norm{\sigma(\epsilon)}^2_0}} \braket{p_1,\ldots,p_n; 0 | \psi(\epsilon)}
\ee
holds for all \(\ket{\psi} \in \ms D_\Delta\), where \(\sigma_{AB}(p;\epsilon)\) is the trucation of the shear. In the last line we have used \(A_0 = 1\) for the invariant vacuum \(\ket{\s_0}\) (\cref{rem:zero-memory-simplify}), we get \(A_{\nu(\epsilon)} = e^{-\tfrac{1}{128\pi^2} \norm{\sigma(\epsilon)}^2_0} \neq 0\) for \(\epsilon > 0\) (from \cref{eq:coherent-constant-change}). Now we obtain the sought for formula for the inner product on \(\Fock_\Delta\).

\begin{thm}[Inner product on \(\Fock_\Delta\) in terms of BMS particles]\label{thm:inp-memory}
The inner product on the memory Fock spaces \(\Fock_\Delta\), restricted to the dense subspace \(\ms D_\Delta\) is given by
\be\label{eq:inp-memory}
    \braket{\psi' | \psi}_\Delta = \lim_{\epsilon \to 0}  \frac{e^{-\tfrac{1}{64\pi^2} \norm{\sigma(\epsilon)}_0^2}}{\abs{A_\nu}^2} \sum_{n=0}^\infty \tfrac{1}{(16\pi)^n} \int_{C_+^n} \lb( \prod_{i=1}^n  d^3p_i \omega_i^{-1} \theta_\epsilon(\omega_i) \rb) \braket{\psi' | p_1,\ldots,p_n;\Delta} \braket{p_1,\ldots,p_n;\Delta | \psi}
\ee
for all \(\ket{\psi}, \ket{\psi'} \in \ms D_\Delta\). In the above formula, the shear \(\sigma_{AB}\) and the corresponding News \(\nu_{AB}\) have memory \(\Delta_{AB}\), \(A_\nu = \braket{\ground;\Delta | \s_\nu} \neq 0\) (as in \cref{eq:0-plane-wave-defn}), and \(\sigma_{AB}(\epsilon)\) is the truncation of the shear as explained above. Further, the formula \cref{eq:inp-memory} is independent of the chosen vacuum \(\ket{\s_\nu} \in \Fock_\Delta\).
\begin{proof}
First we prove the final claim that the formula \cref{eq:inp-memory} is independent of the choice of vacuum \(\ket{\s_\nu}\). Let \(\ket{\s_{\nu'}}\) be another choice of vacuum with News tensor \(\nu'_{AB}\) which has the same memory \(\Delta_{AB}\). The the corresponding shears are related by \(\sigma'_{AB} = \sigma_{AB} + s_{AB}\) where \(s_{AB}\) is a test tensor with zero memory. By the arguments already given in remark \ref{rem:change-vacuum} the overlap of the BMS particle states is invariant under the choice of vacuum state in \(\Fock_\Delta\); thus we only need to consider the prefactor in \cref{eq:inp-memory}. We compute the ratio of the prefactors with different choice of vacua using \cref{eq:coherent-constant-change}
\be\label{eq:inp-prefactor}
    \lim_{\epsilon \to 0} \abs{\tfrac{A_{\nu'}}{A_\nu}}^2 e^{\tfrac{1}{64\pi^2} \lb(\norm{\sigma'(\epsilon)}_0^2 - \norm{\sigma(\epsilon)}_0^2\rb)}
    & = \lim_{\epsilon \to 0} e^{ \lb( \tfrac{1}{4\pi} \Im \nu(\bar s) -\tfrac{1}{64\pi^2} \norm{s}_0^2 - \tfrac{1}{64\pi^2} \norm{\sigma(\epsilon)}_0^2 + \tfrac{1}{64\pi^2} \norm{\sigma'(\epsilon)}_0^2 \rb) } \\
    & = \lim_{\epsilon \to 0} e^{ \lb( \tfrac{1}{4\pi} \Im \nu(\bar s (1-\theta_\epsilon)) -\tfrac{1}{64\pi^2} \norm{s (1-\theta_\epsilon)}_0^2 \rb) } \\
    & = 1
\ee
where the second line uses the identity
\be
    - \norm{\sigma(\epsilon)}_0^2 + \norm{(\sigma + s)\theta_\epsilon}_0^2 &= \norm{s \theta_\epsilon}_0^2 + 2 \Re \braket{s | \sigma(\epsilon)}_0 = \norm{s \theta_\epsilon}_0^2 - 16\pi \Im \nu(\bar s \theta_\epsilon)
\ee
Thus, the inner product formula is independent of the choice of vacuum state.

Now, we prove that this formula is the inner product on \(\Fock_\Delta\). Let \(\op O(\psi)\) be the smeared sum over product of creation operators that maps the chosen vacuum \(\ket{\s_\nu}\) to the state \(\ket{\psi} = \op O(\psi) \ket{\s_\nu} \in \ms D_\Delta\) according to \cref{eq:nice-state}, and similarly, \(\ket{\psi(\epsilon)} = \op O(\psi) \ket{\omega_{\nu(\epsilon)}} \in \Fock_0\) be the corresponding truncated state in \(\Fock_0\) as explained above. Then from \cref{eq:correlation-lim} we have
\be
    \braket{\s_\nu | \op O(\psi')^* \op O(\psi) | \s_\nu}_\Delta &= \lim_{\epsilon \to 0} \braket{\omega_{\nu(\epsilon)} |\op O(\psi')^* \op O(\psi) | \omega_{\nu(\epsilon)} }_0 \\
    \braket{\psi' | \psi}_\Delta &= \lim_{\epsilon \to 0} \braket{\psi'(\epsilon) | \psi(\epsilon)}_0
\ee
Now we write the inner product on the right-hand side in terms of the zero memory \(n\)-particle states, as in \cref{eq:inp-0}, to get
\be
    \braket{\psi' | \psi}_\Delta = \lim_{\epsilon \to 0}  \sum_{n=0}^\infty (16\pi)^{-n} \int_{C_+^n} \lb( \prod_{i=1}^n  d^3p_i \omega_i^{-1} \rb) \braket{\psi'(\epsilon) | p_1,\ldots,p_n; 0} \braket{p_1,\ldots,p_n; 0 | \psi(\epsilon)}
\ee
Finally, we use \cref{eq:pn-trunc} to replace the zero memory \(n\)-particle states with the BMS \(n\)-particle states with memory which proves \cref{eq:inp-memory}.
\end{proof}
\end{thm}
When the memory vanishes we can take the limit inside the sum in \cref{eq:inp-memory} since each term has a finite limit, to get the zero-memory inner product \cref{eq:inp-0}. But for non-zero memory, \(\Delta_{AB} \neq 0\), we must take the limit last since the individual integrals in the formula will not converge in the limit. In fact, the integrals along with the infinite sum together give an overall factor which precisely cancels the exponential factor in \cref{eq:inp-memory}, so that the final expression converges as \(\epsilon \to 0\).

Next, we give a plausible interpretation of \cref{eq:inp-memory} in terms of a direct integral. Define \(C_+^0 = \set{0}\) as the one element set and take the set \(\mf C_+\) defined as the disjoint union
\be\label{eq:base-DI}
    \mf C_+ \defn \bigsqcup_{n=0}^\infty C_+^n = \set{(p_1,\ldots,p_n) : n \geq 0 \text{ and } (p_1,\ldots,p_n) \in C_+^n }
\ee
At each point \((p_1,\ldots,p_n)\) of \(\mf C_+\), let  \(\Hilb_{p_1,\ldots,p_n}\) be the Hilbert space of complex-valued test tensors \(s_{A_1B_1\ldots A_nB_n}\) which are symmetric and traceless in each pair of indices \(A_iB_i\) and totally symmetric with respect to interchange of any such pair with another. The inner product on \(\Hilb_{p_1,\ldots,p_n}\) is simply the contraction of their indices
\be
    {s'}^{A_1B_1\ldots A_nB_n} s_{A_1B_1\ldots A_nB_n}
\ee
Now, for \(\epsilon > 0\), consider the family of measures \(\mu_{\Delta,\epsilon}((p_1,\ldots,p_n))\) on \(\mf C_+\) defined by
\be
    d\mu_{\Delta,\epsilon}((p_1,\ldots,p_n)) \defn \frac{e^{-\tfrac{1}{64\pi^2} \norm{\sigma(\epsilon)}^2_0}}{\abs{A_\nu}^2} (16\pi)^{-n}~ \delta_n~ \prod_{i=1}^n d^3 p_i \omega_i^{-1} \theta_\epsilon(\omega_i)
\ee
where \(\nu_{AB}(p)\) is a News tensor with memory \(\Delta_{AB}\), \(\sigma_{AB}\) is the corresponding shear, and \(\sigma_{AB}(\epsilon)\) is the truncated shear are explained above. The \(\delta_n\) is the discrete Dirac measure supported at \(n\). The same computation as in \cref{eq:inp-prefactor} above, without taking the final limit \(\epsilon \to 0\), shows that the measures defined with different News tensors with the same memory are equivalent; hence we have labeled the measure with the memory \(\Delta\) instead of the News tensor. Now we construct a direct integral Hilbert space over \(\mf C_+\) by integrating the Hilbert spaces \(\Hilb_{p_1,\ldots,p_n}\) with the above family of measures,
\be\label{eq:DI}
    \Fock_{\Delta,\epsilon} \defn \int_{\mf C_+}^\oplus d\mu_{\Delta,\epsilon} \Hilb_{p_1,\ldots,p_n}
\ee 
A state in \(\Fock_{\Delta, \epsilon}\) is then represented as a function \(\psi\) on \(\mf C_+\) such that at \((p_1,\ldots,p_n)\) we have
\be\label{eq:state-DI}
    \psi((p_1,\ldots,p_n)) \defn \braket{p_1,\ldots,p_n; \Delta | \psi} \in \Hilb_{p_1,\ldots,p_n}
\ee
where \(\ket{\psi} \in \Fock_\Delta\) and two functions are equivalent if they differ only on a set of measure zero with respect to \(\mu_{\Delta,\epsilon}\).

Now, when \(\Delta = 0\), the limit of the measures as \(\epsilon \to 0\) can be taken and the limiting measure is
\be\label{eq:mu-DI-0}
    d\mu_0((p_1,\ldots,p_n)) = \lim_{\epsilon \to 0}d\mu_{0,\epsilon}((p_1,\ldots,p_n)) = (16\pi)^{-n}~ \delta_n~ \prod_{i=1}^n d^3 p_i \omega_i^{-1} 
\ee
Then, the direct integral above gives the usual direct sum structure of the zero memory Fock space in terms of \(n\)-particle states (\cref{eq:fock-0})
\be
    \Fock_0 \,\cong\, \int_{\mf C_+}^\oplus d\mu_0((p_1,\ldots,p_n)) \Hilb_{p_1,\ldots,p_n} \,\cong\, \bigoplus_{n\geq 0} \lb[ (16\pi)^{-n} \int_{C_+^n} \lb(\prod_{i=1}^n d^3 p_i \omega_i^{-1} \rb)~\Hilb_{p_1,\ldots,p_n} \rb]
\ee
In fact, for any memory the direct integral \cref{eq:DI} is unitarily equivalent to the zero memory Fock space when \(\epsilon > 0\); this is just a restatement of our truncation procedure described above. However for any non-zero memory, corollary \ref{cor:infinite-bms-particles} implies that any subset of \(\mf C_+\) containing only finitely-many BMS particle numbers must have measure zero, and thus the direct sum structure of \(\Fock_0\) cannot be generalized to \(\Fock_\Delta\) in terms of the BMS particles. We conjecture that for \(\Delta \neq 0\) the limit of the measures \(\mu_{\Delta,\epsilon}\) as \(\epsilon \to 0\) exists as a measure \(\mu_\Delta\) on \(\mf C_+\). Since, the sets of measure zero for \(\mu_\Delta\) and \(\mu_0\) do not agree, the limiting measure \(\mu_\Delta\) must be inequivalent to \(\mu_{0}\). If this conjectured limit measure exists, the inner product formula \cref{eq:inp-memory} can be given a precise measure theoretic meaning and the memory Fock spaces can be given the structure of a direct integral
\be
    \ms F_\Delta \,\cong\, \int_{\mf C_+}^\oplus d\mu_\Delta((p_1,\ldots,p_n)) \Hilb_{p_1,\ldots,p_n}
\ee
However, proving the existence of such a limit of measures would require considerable mathematical machinery, so we leave the existence of \(\mu_\Delta\) as purely conjecture.


\bibliographystyle{JHEP}
\bibliography{asymp-quantization}      
\end{document}